\documentclass[a4paper,pra,twocolumn,superscriptaddress]{revtex4-1}

\usepackage{graphicx}
\usepackage{amsmath}
\usepackage{epstopdf}
\newcommand{\uvec}{\boldsymbol{u}}
\newcommand{\kvec}{\boldsymbol{k}}
\newcommand{\Rvec}{\boldsymbol{R}}
\newcommand{\rvec}{\boldsymbol{r}}
\newcommand{\rveci}{\boldsymbol{r}_{i}}

\newcommand{\Dvec}{\boldsymbol{D}}

\newcommand{\zetavec}{\boldsymbol{\zeta}}
\newcommand{\drm}{{\rm d}}
\newcommand{\tauvec}{\boldsymbol{\tau}}
\newcommand{\vacu}{|0\rangle}
\newcommand{\Adag}{A^{\dagger}}
\newcommand{\Bdag}{B^{\dagger}}
\newcommand{\Cdag}{C^{\dagger}}

\begin{document}

\author{J.P. Hague}
\affiliation{School of Physical Sciences, The Open University, Milton Keynes, MK7 6AA, UK}

\author{P.E. Kornilovitch}
\affiliation{Department of Physics, Oregon State University, Corvallis, OR, 97331, USA}

\author{C. MacCormick}
\affiliation{School of Physical Sciences, The Open University, Milton Keynes, MK7 6AA, UK}  


\title{A cold atom quantum simulator to explore pairing, condensation, and pseudogaps in extended Hubbard--Holstein models}

\begin{abstract} 
We describe a quantum simulator for the Hubbard--Holstein model (HHM), comprising two dressed Rydberg atom species held in a monolayer by independent painted potentials, predicting that boson-mediated preformed pairing, and Berezinskii--Kosterlitz--Thouless (BKT) transition temperatures are experimentally accessible. The HHM is important for modeling the essential physics of unconventional superconductors. Experimentally realizable quantum simulators for HHMs are needed: (1) since HHMs are difficult to solve numerically and analytically, (2) to explore how competition between electron-phonon interactions and strong repulsion affects pairing in unconventional superconductors, (3) to understand the role of boson-mediated local pairing in pseudogaps and fermion condensates. We propose and study a quantum simulator for the HHM, using optical lattices, painted using zeros in the AC stark shift, to control two Rydberg atom species independently within a monolayer. We predict that interactions are sufficiently tunable to probe: (1) both HHMs and highly unconventional phonon-mediated repulsions, (2) the competition between intermediate-strength phonon and Coulomb mediated interactions, (3) BKT transitions, and preformed pairing that could be used to examine key hypotheses related to the pseudogap. We discuss how the quantum simulator can be used to  investigate boson-mediated pairing and condensation of fermions in unconventional superconductors.

\end{abstract}

\maketitle

\section{Introduction}

Boson-mediated pairing of fermions has not yet been observed in cold atom experiments. Cold atom quantum simulators have been very successful for simulating Hubbard models. The Mott metal-insulator and superfluid-insulator transitions have both been observed \cite{bloch2008a, bloch2012a}. The Feshbach resonance can be tuned into an attractive regime, allowing local pairing in attractive Hubbard models to be measured directly using gas microscopy \cite{mitra2018a}. In solid state systems, attractive Hubbard models are the effective Hamiltonian arising from boson-mediated interactions, so it would be of significant interest to probe such interactions directly. 

Probing boson-mediated pairing in a quantum simulator is technically demanding, but potentially highly rewarding as this pairing reflects the mechanism of many superconductors. The recent discovery of hydrogen based superconductors at ambient temperatures makes boson-mediated superconductivity particularly pertinent \cite{drozdov2015,drozdov2019}.

Unconventional superconductors often contain significant Coulomb repulsions and boson-mediated couplings, e.g. electron-phonon interactions \cite{bcs, zhang1988a, zhao1996a, lanzara2001a, song2019a, li2019a}. Furthermore, recent exact numerics  provide strong upper bounds on superconductivity in the popular Hubbard model \cite{qin2019a}, identifying the need to include additional interactions alongside this model to explain superconductivity in such materials.

The HHM and its extensions \cite{hubbard1964a,holstein1959a} contain the essence of these interactions, but the HHM lacks reliable numerical and analytical solutions. A tunable quantum simulator would allow this model to be explored without the complications associated with the multitude of competing interactions and phases found in unconventional superconductors.  Moreover, the phase diagram, including the onset of superconductivity, the opening of gaps and pseudogaps, the Bardeen--Cooper--Schrieffer (BCS) to Bose--Einstein condensate (BEC) crossover (from the point of view of boson mediated pairing), and the Mott insulating state, could be observed directly by tuning the interaction strength without the limitations of stoichiometry and pressure. While some of these phenomena have been observed individually in purely fermionic quantum simulators, the interplay between these phases and boson mediated interactions has not been measured.

The innovations within the quantum simulator for the HHM proposed here are: (1) exploitation of zeros in the AC stark shift to generate bipartite lattices within a single optical pancake to reduce experimental complexity, (2) use of Rydberg mediated interactions to tune electron-phonon interaction and Coulomb repulsion independently, (3) the possibility to investigate highly unconventional repulsive interactions mediated via phonons, (4) the possibility to explore boson-mediated pairing, (5) the possibility to examine pseudogap physics, (6) the potential to investigate the BKT transition.





This paper is organised as follows: Section \ref{sec:qs} introduces the proposed experimental setup of the quantum simulator. In Section \ref{sec:hamiltonian} we derive the Hamiltonian of the simulator. In Section \ref{sec:phase} we discuss the phase diagram of the simulator in the limits of strong coupling and high phonon frequency. In Section \ref{sec:conclusions} we discuss the how the phase diagram of the quantum simulator could be used to examine the properties of unconventional superconductors. We also include an Appendix with the full mathematical details of the phase diagram calculations.

\section{Quantum simulator}
\label{sec:qs}

In this section, we describe how two atomic species can be trapped within different optical lattices, and describe the patterns of optical lattice potentials that can lead to a quantum simulator for a Hubbard--Holstein model. Throughout this paper, we will work with dressed Rydberg atoms since they provide long range interactions.


\subsection{Form of the optical lattice}

The optical lattice contains a single optical pancake with laser wavelength $\lambda_{\rm pan}$ and width $w_{\rm pan}$, within which potentials are painted using Gaussian beams \cite{henderson2009a}.  The total lattice potential is 
\begin{equation}
    V(\Rvec) = \frac{1}{N_{S}}\sum_{i}\sum_{j=1}^{N_{S}}V_{\rm spot,i}(\rvec-\rvec_{i}-\Dvec_{j})+V_{\rm pan}(z),
\end{equation}
where 
\begin{equation}
    V_{\rm pan}(z) = -V_{0,\rm pan} \exp(-2z^2/w_{\rm pan}^2)
\end{equation}
$\rvec$ is a vector that lies within the plane of the optical pancake, $\Rvec=\rvec+z\kvec$, $\rvec_{i}$ represent the center of a spot arrangement, $\Dvec_{j}$ are basis vectors, $N_{S}$ are the number of spots forming the basis, and the $z$-axis is perpendicular to the pancake. In the fermion lattice $\Dvec_{j}$ are always zero, and $N_{S}$ is always one.

The spot potential has the form, 
\begin{equation}
V_{\rm spot}(r)=-V_{0}\exp(-2r^2/w^2).
\end{equation}
Fermion beams have waist,  $w_{\rm f}$, and phonon beams have waist $w_{\rm ph}$.

\begin{figure}
\includegraphics[width=0.45\textwidth]{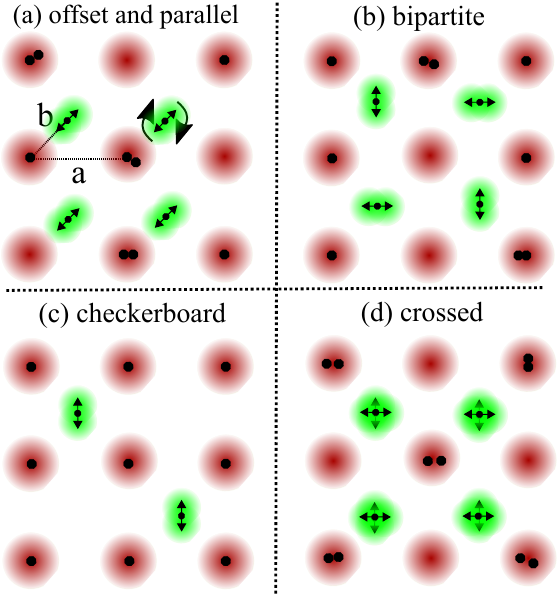}
\caption{[Color online] Spot arrangements considered in this paper. Dark gray [red] spots that are not associated with arrows trap fermions. Black dots without arrows represent fermions. Two overlapping light gray [green] spots with arrows trap atoms (which would normally be bosons), the vibrations of which represent phonons. The bosons sites are represented by black dots with arrows. Arrows represent the polarization of phonon modes. Phonon spot arrangements can be translated and rotated relative to the fermion lattice (see panel (a)). 
} 
\label{fig:patterns}
\end{figure}



 The key feature of this quantum simulator is that the optical lattice has a basis of two spot types, that independently trap different atomic species (Fig. \ref{fig:patterns}).  Simulator properties are fixed by the pattern of optical lattice potentials. The lattice constant is $a$.

One lattice contains atoms in a Mott insulating state, that represent the nuclei in condensed matter systems, which are able to vibrate to represent phonons, but are not able to hop between lattice sites. The motion of the atoms in this lattice is represented in Fig. \ref{fig:patterns} using small arrows associated with black dots, and the spot shading is light grey (green). The phonon lattice should be kept in a Mott insulating state to ensure that there is a single atom per site, just as there is a single nucleus per atom in a condensed matter system. 

Each site in the phonon lattice consists of multiple spots with a separation close to the Raleigh limit. This allows the lattice to be deep, and yet provides broad sites within which the atoms can oscillate. This will be discussed in more detail later in this paper, when the Hamiltonian corresponding to the quantum simulator is derived.

The second lattice contains itinerant fermions that represent electrons. The spots in this lattice are shaded dark gray (red). Fermions in this lattice are represented by dots. The fermion lattice may be partially filled and fermions may hop between sites. There may be 0,1 or 2 fermions per lattice site.

We investigate models generated by several spot configurations, which are shown schematically in Fig. \ref{fig:patterns}. Rotating phonon spot patterns with respect to the fermion lattice can change model properties. The phonon lattice can be offset from the midpoints between the fermion lattice sites by changing the distance, $b$, and this can also modify the properties of the quantum simulator. The effect of these changes will be discussed later in the paper. 
 
  We note that vibrating fermions could also be used to represent phonons. However, we do not consider this here since  the presence of bosonic atoms allows for more straightforward setup of the system: the singly occupied phonon sites can be produced from a Bose-Einstein condensate via a superfluid-Mott insulator technique (this method ensures single atom occupation of the sites).
 
 
 The components of the quantum simulator and their condensed matter analogues are summarized in Table~\ref{tab:qs}.

\subsection{Species dependent optical lattices}

Two atomic species, a fermion representing electrons and a boson that can vibrate to represent phonons, can be trapped in different, but coexisting, lattices by exploiting \emph{state dependence in the AC stark shift} \cite{mandel2003,schrader2001}. In the following, we consider bosonic $^{87}$Rb and fermionic $^{40}$K, trapped  by linearly polarized lasers of different wavelengths $\lambda_{\rm ph}$ and $\lambda_{\rm f}$ respectively. 

In general, a laser blue detuned from a given transition induces an atomic dipole moment $\mathbf{d}_{\mathrm{A}}$ oscillating in anti-phase to (and hence anti-aligned with) the laser's electric field $\mathbf{E}_\mathrm{L}$; the potential energy $U_\mathrm{A}=-\mathbf{d}_{\mathrm{A}} \cdot \mathbf{E}_{\mathrm{L}} > 0$. On the other hand, when the laser is red detuned from a given transition, the induced atomic dipole oscillates in phase with the laser's electric field and the potential energy of the atom is $U_\mathrm{A}<0$.


The physical origins of atomic-species-dependent potentials are cancellations between these potential energies that occur when radiation is detuned from two, closely-separated, transition lines (for example the D1 and D2 transition lines), such that radiation with a wavelength between the lines is red detuned relative to one line, and blue detuned relative to the other. These zeros in the AC stark shift lead to a powerful scheme for trapping single atomic species. Alkali atoms can be optically trapped by lasers detuned from the strong D1 and D2 transitions that couple the $nS_{1/2}$ ground state to the $nP_J$ states, where $J=1/2$ for the D1 transition line and $J=3/2$ for the D2 transition. We represent the detuning from the transition as $\delta_i=\omega_i-\bar{\omega}_\mathrm{Las}$ where $i$ is either 1 or 2 for the D1 or D2 lines, $\omega_i$ are the frequencies of the D1 or D2 transition, and $\bar{\omega}_\mathrm{Las}$ is the laser frequency. In the case of the large detunings used in optical traps, the potential energy of a ground state alkali atom with total angular momentum $F$, bathed in light with an intensity, $I(\mathbf{r})_\mathrm{Las}$, is
\begin{eqnarray}
    V(\mathbf{r})_\textrm{trap}& = &\frac{\hbar I(\mathbf{r})_\mathrm{Las}}{24 I_{\textrm{Sat}}}\left(\left(\frac{\Gamma_{1}^2}{ \delta_{1}}+2\frac{\Gamma_{2}^2}{ \delta_{2}}\right)\right.\nonumber\\
    & &\left. -g_F m_F \sqrt{1-\epsilon ^2}\left(\frac{\Gamma_{1}^2}{\delta_{1}}-\frac{\Gamma_{2}^2}{ \delta_{2}}\right)\right),
    \label{eqn:trapDepth}
\end{eqnarray}
where $m_F$ is the magnetic quantum number of the atom, and $g_F$ is the corresponding Land\'e g-factor. 
The polarisation vector of the laser beam is $\hat{\epsilon} = (\sqrt{1+\epsilon}\hat{x}+i\sqrt{1-\epsilon}\hat{y})/\sqrt2$ where $\epsilon$ is the ellipticity. In this work we choose linear polarised light, where $\epsilon=0$, which ensures that the two fermionic spin states experience the same potential energy. 

The properties of D1 and D2 transitions in both $^{40}$K and $^{87}$Rb are well established. For the $^{40}$K, D1 and D2 transitions, the saturation intensity is $I_\mathrm{Sat} = 17.5 \;\mathrm{W \; m}^{-2}$, the D1 transition wavelength and linewidth is $\lambda = 770.1$ nm and $\Gamma_1 = 2\pi\times 5.95 \;$MHz respectively; the D2 transition wavelength and linewidth is $\lambda = 766.7$ nm,  and $\Gamma_1 = 2\pi\times 6.03 \;$MHz respectively.
For $^{87}$Rb, the saturation intensity is $I_\mathrm{Sat} = 16.7 \;\mathrm{W \; m}^{-2}$, the D1 transition wavelength and linewidth is $\lambda = 795.0$ nm and $\Gamma_1 = 2\pi\times 5.74 \;$MHz respectively; the D2 transition wavelength and linewidth is $\lambda = 780.2$ nm, and $\Gamma_2 = 2\pi\times 6.06 \;$MHz respectively.  

\begin{figure}
\includegraphics[width=0.45\textwidth]{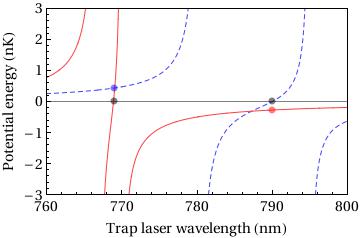}
\caption{[Color online] The potential energies of the fermionic $^{40}$K (solid red line) and bosonic $^{87}$Rb atoms (blue dashed line) according to Eq.~(\ref{eqn:trapDepth}) is shown as a function of trap laser wavelength, taking the prefactor $\hbar I_\mathrm{Las}/24I_\mathrm{Sat}=1\; \mathrm{nK}$. When illuminated by a laser tuned to 768.97 nm, the potential energy of the $^{40}$K atoms vanishes (indicated by a black point) but that of the $^{87}$Rb atoms is positive (indicated by the gray [blue] point on the dashed curve). Similarly, when a laser is tuned to 790.07 nm, the $^{40}$K atoms experience a negative potential energy (indicated by the gray [red] point on the solid curve) but the potential energy of the $^{87}$Rb  atoms vanishes. Exploiting these conditions, a two color optical trapping setup can trap $^{40}$K and $^{87}$Rb  atoms in mutually exclusive trapping potentials.} 
\label{fig:traps}
\end{figure}

Specific wavelengths for the lasers, $\lambda_{\rm f}$ and $\lambda_{\rm ph}$ can be chosen using Eq.~(\ref{eqn:trapDepth}). The approach to selecting $\lambda_{\rm f}$ and $\lambda_{\rm ph}$ is demonstrated in Fig.~\ref{fig:traps}, which shows Eq.~(\ref{eqn:trapDepth}) plotted as a function of trap laser wavelength for each atom, taking the prefactor $\hbar I_\mathrm{Las}/24I_\mathrm{Sat}=1\; \mathrm{nK}$. It can be seen that the potential energy experienced by $^{40}$K is zero when the laser is tuned between the $^{40}$K D1 and D2 transitions at 768.97 nm, whereas $^{87}$Rb atoms experience a positive potential energy (proportional to the laser intensity). The zero in potential energy arises here because the laser is red detuned from the D2 transition and blue detuned from the D1 transition, such that induced atomic dipole moment is canceled. A similar situation occurs when the laser is tuned between the D1 and D2 lines of $^{87}$Rb at $\lambda=790.07$ nm, except that in that case it is the $^{40}$K atoms experience a negative potential energy and the $^{87}$Rb atoms experience zero potential energy. Thus, lattice potentials are blue (red) detuned for $^{87}$Rb ($^{40}$K), so bosonic atoms are trapped in an ``inverse'' lattice where absence of light leads to confinement. For convenience, we discuss attractive potentials for both species,  but these can easily be painted from repulsive ones.

We note that the approach of exploiting zeros in the AC stark shift is flexible - one could choose to work with atoms other than $^{87}$Rb, e.g. $^{133}$Cs, which could be trapped using a laser blue detuned at the $^{40}$K wavelength $\lambda_{\mathrm{40}}\simeq 769$ nm trap and fermionic $^{40}$K atoms trapped in a red detuned using the $^{133}$Cs wavelength of zero AC stark shift $\lambda_\mathrm{133}\simeq 866.4$ nm.

 \section{Hamiltonian}
 \label{sec:hamiltonian}

\subsection{Phonons}

\begin{table*}
  \caption{Summary of the quantum simulator and correspondence with condensed matter systems.}
  \begin{tabular}{|c|c|c|}
  \hline
 &   {\bf quantum simulator} & {\bf condensed matter} \\
    \hline
  \textit{\textbf{fermion}} & fermionic $^{40}$K & electron \\
  \textit{\textbf{lattice potential}} & single spot potential & nuclear potential\\
\textit{\textbf{phonon}} &    $^{87}$Rb oscillations in multi-spot potential & nuclear vibrations\\
\textit{\textbf{fermion-phonon interaction}} &    Rydberg-phonon interaction & electron-phonon interaction\\
\textit{\textbf{Hubbard U}} &    Feshbach resonance & Coulomb repulsion\\
\hline
    \end{tabular}
    \label{tab:qs}
  \end{table*}

\begin{figure}
    \centering
    \includegraphics[width = 0.48\textwidth]{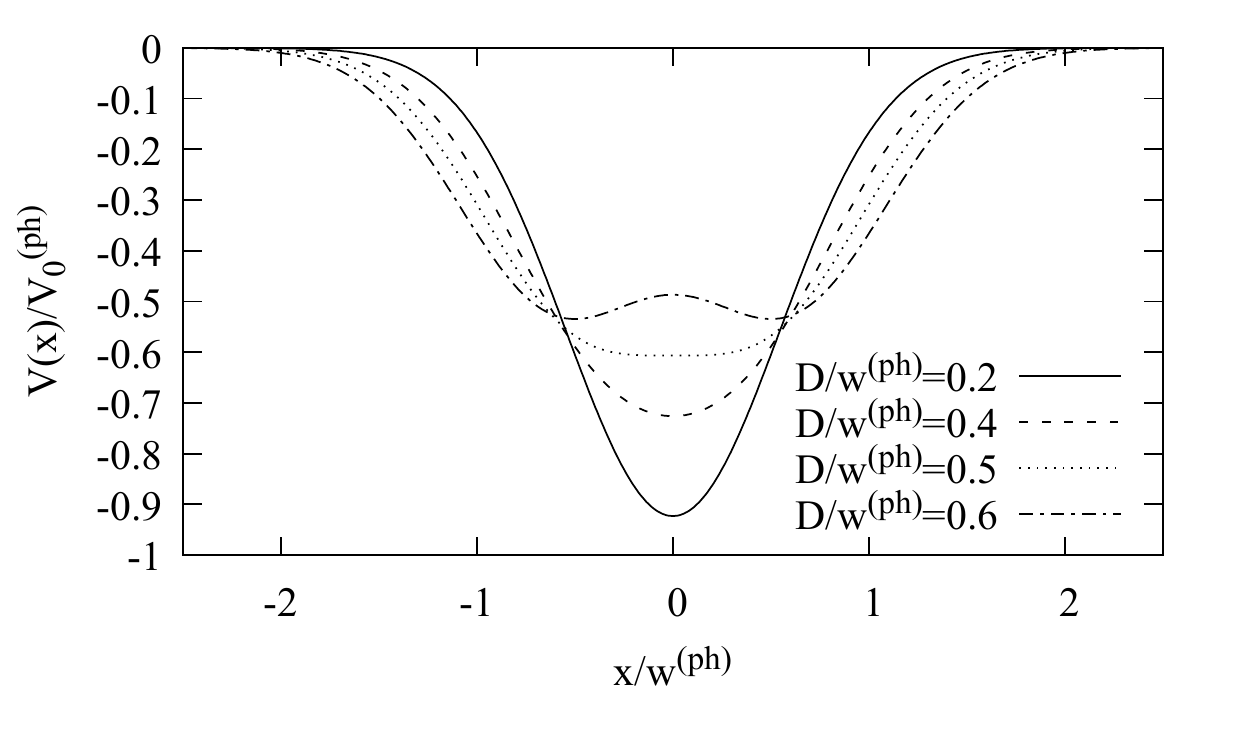}
    \caption{By painting two spots a distance $2D$ apart, the curvature of the origin at the minimum can be controlled ranging from completely flat for $D=w^{\rm (ph)}/2$ to the curvature for a single potential for $D=0$. Thus the phonon frequency can be reduced by an order of magnitude. Note that $D\leq w^{\rm (ph)}/2$ should be selected so that a double well potential does not form.}
    \label{fig:twospot}
\end{figure}

In the quantum simulator, vibrations of the $^{87}$Rb atoms in multi-spot patterns are used to represent phonons in a condensed matter system. In this section, we briefly summarize the phonon subsystem of the quantum simulator, and explain why multi-spot patterns are needed. Further information on multi-spot patterns can be found in Ref.~[\onlinecite{hague2012a}].

Condensed matter systems have a nucleus per atom, the vibrations of
which are phonons. Thus we require that there is a single $^{87}$Rb
atom per site in the quantum simulator to match the situation in the
condensed matter system. One way to achieve this is to put the
$^{87}$Rb bosons representing phonons into a Mott insulating
state. This can be achieved by making the $^{87}$Rb optical lattice
deep.

The energy scales of phonons in a condensed matter system are
typically 1 or 2 orders of magnitude smaller ($\sim 10-100$ meV) than
the energy scales of electrons ($\sim 1$ eV). We require that the
relative energy scales of hopping and phonons in the quantum simulator
follow a similar hierarchy. This means that the potential at the bottom
of the well of the phonon sites must be slowly varying. 

In order to ensure similar hierarchy of energy scales in a quantum
simulator, the atoms that represent phonons (e.g. $^{87}$Rb) must
oscillate in a deep, yet broad, trap with small frequencies.
Generating a Mott insulating state and small energy scales (and thus
frequencies) for phonons presents the following challenge: A deep trap
is needed to generate the Mott insulating state; yet the deeper the
trap, the higher the phonon frequencies. Painted potentials offer a solution to this apparent contradiction.


A broad and deep trap for phonons can be painted using several closely
positioned spots. Multi-spot arrangements have an effective potential,
\begin{equation}
V_{\rm ph}(\rvec) = \frac{1}{N_S}\sum_{i}\sum_{j=1}^{N_S} V_{\rm spot}(\rvec-\rvec_{i}-\Dvec_{j})
\end{equation}
where $\Dvec_{j}$ are the displacements of the phonon spots from the mean position $\rvec_{i}$ \cite{hague2012a}, and $N_S$ is the number of spots forming the phonon site. 

This potential is shown for two spots in Fig. \ref{fig:twospot}. If the spots are spaced around the full-width half-maximum distance of the Gaussian beam, then the second derivative of the potential where the spots meet can be significantly reduced, thus reducing the frequency of oscillations in the trap.  A benefit of the painted potential approach to making this kind of potential is that the laser intensity required to paint a multi-spot arrangement with the same central depth, but lower phonon frequency, is approximately the same as the intensity required to paint a single spot. Thus, large lattices with the multi-spot basis can be formed.

Phonon properties can be derived from the dynamical matrix, $A^{ij} = \partial^2 V_{\rm ph}/\partial u_{i}\partial u_{j}|_{0}$, where $u_{i}$ is the atom displacement.  Eigenvalues of the matrix are $\omega_{ph,\nu}$ and eigenvectors define the phonon polarization, $\zetavec_{\nu}$. For phonon spots separated by a distance $2D$ on a single axis there is a single polarization with frequency 
\begin{equation}
\omega_{\rm ph} = 2 \exp\left[-\frac{D^2}{w_{\rm ph}^2}\right]  \sqrt{\frac{V_{\rm 0,ph}(w_{\rm ph}^2-(2D)^2)}{M_{\rm Rb} w_{\rm ph}^{4}}}.
\end{equation}
Similar frequencies will be found for four spot arrangements \footnote{Note a difference in the definition of the spot potentials compared to Ref. \cite{hague2012a}, which leads to a slight difference in the expression.}.

The phonon contribution to the Hamiltonian is $H_{ph} =
\sum_{\nu,i}\hbar\omega_{{\rm ph},\nu} d^{\dagger}_{i\nu}
d_{i\nu}$. $d^{\dagger}$ creates a phonon and $\omega_{\rm ph,\nu}$ is
the phonon frequency of mode $\nu$. 
For two-dimensional spot arrangements there are two polarizations. Phonons are not coupled between sites, so are $\kvec$ independent.


\subsection{Rydberg-phonon interactions}

The Rydberg-phonon interaction in the quantum simulator is the analogue of the electron-phonon interaction in a condensed matter system. Rydberg-phonon interaction arises from coupling between dressed Rydberg atoms of different species (i.e. the $^{40}$K and $^{87}$Rb),
\begin{equation}
  V_{R}(r) = \bar{\alpha}^4\frac{C_{6}}{r^6+C_{6}/2\Delta_{\rm 2p}} = \frac{\tilde{V}_{\rm Ryd}}{r_{c}^{\eta}+r^{\eta}},
  \label{eqn:rydberginteraction}
\end{equation}
 Equation~(\ref{eqn:rydberginteraction}) is calculated with van Vleck perturbation theory and is reliable if $\bar{\alpha}= \Omega_{\rm 2p}/2\Delta_{\rm 2p}\lesssim 0.2$ \cite{hague2017a}. Here, we use parameters for Van der Waals Rydberg-Rydberg coupling with $\eta=6$ so that we limit to near-neighbor interactions, but $\eta=3$ is also possible \cite{hague2012a}. Rabi frequency, $\Omega_{\rm 2p}$, characterizes the atom-laser coupling. $\Delta_{\rm 2p}$ is laser detuning from the ground $\rightarrow$ Rydberg state transition. $\tilde{V}_{\rm Ryd}$ could be repulsive or attractive without loss of generality. Equation~(\ref{eqn:rydberginteraction}) can be Taylor expanded, 
 
 \begin{align}
  V_{R}(\rvec+\uvec) & =V_{R}(\rvec)+\uvec\cdot \nabla V_{R}(\rvec) + \cdots\\
  & = V_{R}(\rvec) - \tilde{V}_{\rm Ryd} \eta\uvec\cdot\hat{\rvec} r^{\eta-1}/(r^{\eta}+r_c^{\eta})^2 + \cdots
 \end{align}
 with phonons quantized via, 
 \begin{equation}
  \uvec_{i} = \sum_{\kvec,\nu}\sqrt{\frac{\hbar}{2NM_{\rm Rb}\omega_{{\rm ph}, \kvec\nu}}}\zetavec_{\kvec\nu}(d_{\kvec\nu}e^{-i\kvec\cdot\Rvec_{i}}+d^{\dagger}_{\kvec\nu}e^{i\kvec\cdot\Rvec_{i}}).  
 \end{equation}

Thus, the Rydberg-phonon interaction is described by
\begin{equation}
H_{\rm R-ph} = -\left(\frac{\hbar}{2M_{\rm Rb}\omega_{\rm ph}}\right)^{1/2} \sum_{ij,\nu}f_{ij,\nu}n_{i}(d^{\dagger}_{j\nu}+d_{j\nu})
\end{equation}
where, $\tilde{\rvec}_{ij} = \rvec_{i}-\Rvec_{j}$, and $\hat{\rvec}_{ij} = \tilde{\rvec}_{ij}/|\tilde{r}_{ij}|$. 
\begin{equation}
f_{ij,\nu} = \tilde{V}_{\rm Ryd}\eta\int\drm^3\rvec \phi^{2}_{0}(\rvec)\frac{\zetavec_{\nu,j}\cdot\hat{\rvec}_{ij}|\tilde{\rvec}_{ij}|^{\eta-1}}{\left(|\tilde{\rvec}_{ij}|^{\eta}+r_c^{\eta}\right)^2}.
\end{equation}
$\phi_{0}$ is the ground state harmonic oscillator wave function. For simplicity, we assume that $\phi_{0}(\rvec) = \delta(\rvec)$, which is valid if the typical length scale of the harmonic oscillator wavefunction $\sqrt{\hbar/M_{\rm Rb}\omega_{\rm ph}} \ll a$, where $a$ is lattice spacing. So,
\begin{equation}
f_{ij,\nu} = \tilde{V}_{\rm Ryd}\eta \frac{\zetavec_{\nu,j}\cdot\hat{\rvec}_{ij}|\tilde{\rvec}_{ij}|^{\eta-1}}{\left(|\tilde{\rvec}_{ij}|^{\eta}+r_c^{\eta}\right)^2} \: . 
\end{equation}
%
%

We briefly note that the quality of this Rydberg-phonon interaction term is contingent on sufficiently small oscillations of $^{87}$Rb atoms around their equilibrium positions, and also requires that the typical length scale of the harmonic oscillator wavefunction $\sqrt{\hbar/M_{\rm Rb}\omega_{\rm ph}} \ll a$. The quantum simulator can always be tuned into a state where the harmonic approximation holds. For example, we can control the size of oscillations via the depth of the lattice.

\begin{figure}
\includegraphics[width=0.48\textwidth]{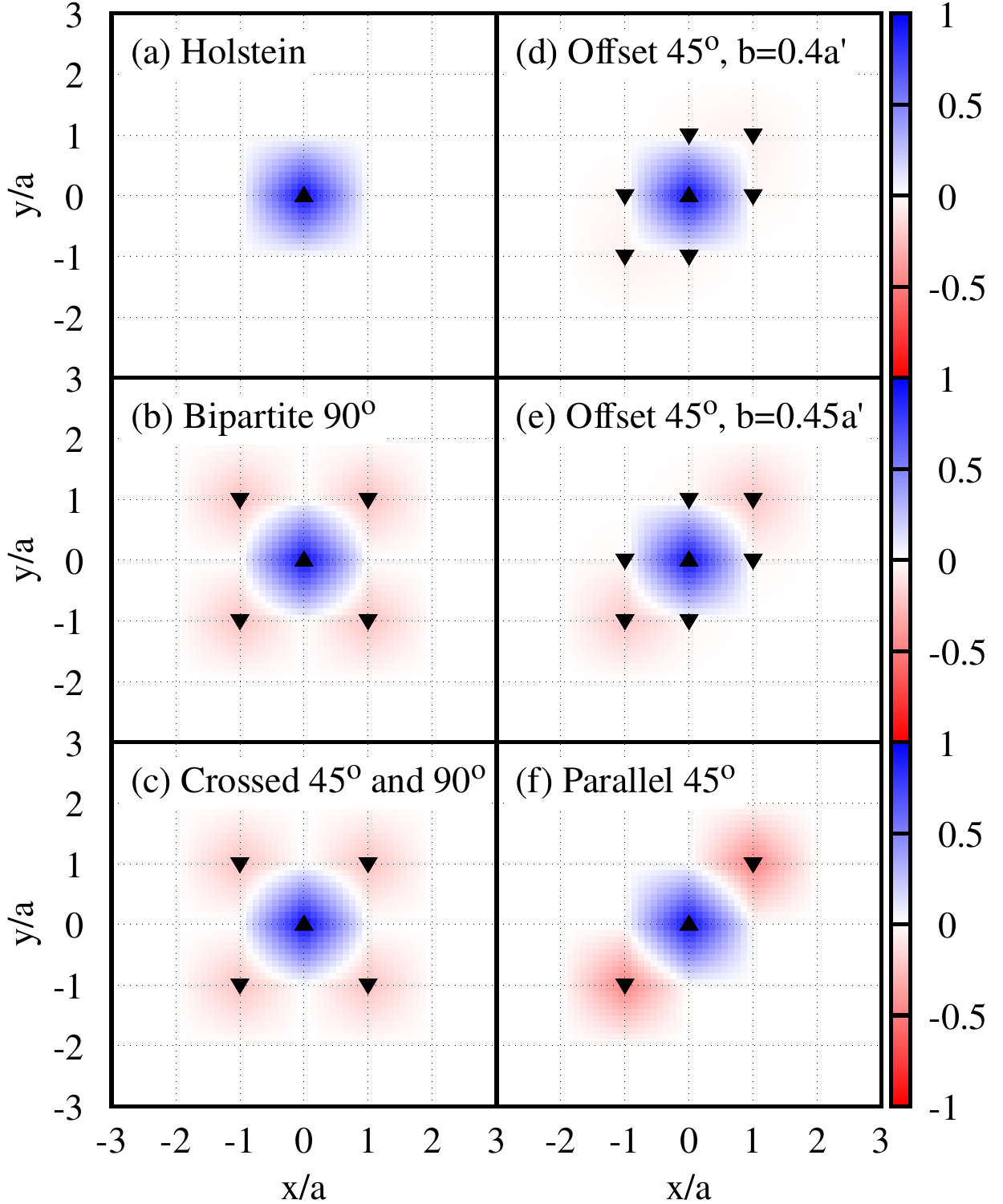}
\caption{[Color online] By changing spot patterns, the effective phonon-mediated interaction ratio, $\Phi_{xy}/\Phi_{00}$, can be controlled and repulsive NN terms can be modified or removed.  The values of $\Phi_{xy}/\Phi_{00}$, resulting from  several phonon spot patterns, are plotted. The ratio $\Phi_{xy}/\Phi_{00}$ is dimensionless, and $\Phi_{xy}$ is defined in Eq.~(\ref{eqn:effectiveinteractionphi}). The lowest off-site repulsion is found for offset phonon positions. Gray shading [blue / red shading] represents the magnitude of $\Phi_{xy}/\Phi_{00}$ with up (down) pointing triangles (shown if $|\Phi_{xy}/\Phi_{00}|>1\%$) represents attraction (repulsion). $r_c = 0.1a$. 
}
\label{fig:phi}
\end{figure}

\subsection{Hopping and Hubbard $U$}
 
Standard forms have been derived for the hopping, $t$ and Hubbard $U$ in Ref.~ [\onlinecite{bloch2008a}]. We summarize them here.
 
In our proposed quantum simulator, $^{40}$K fermions hop between lattice sites according to the term, 
\begin{equation}
H_{\rm hop} = -t\sum_{ij\sigma} c^{\dagger}_{i\sigma} c_{j\sigma},
\end{equation}
where $c^{\dagger}_{i\sigma}$ creates a fermion on site $i$ with spin $\sigma$. The hopping may be approximated as 

\begin{equation}
t\approx 4E_{\rm rec}^{1/4} V_{0}^{3/4} \exp[-2(V_{0}/E_{\rm rec})^{1/2}]/\sqrt{\pi},
\end{equation}
where $E_{\rm rec}=\hbar^2 \pi^2/2M_{\rm K}a^2$ \cite{bloch2008a}. 


The final term in the Hamiltonian is the Hubbard interaction, 
\begin{equation}
H_{\rm Hub}=U_{\rm Fesh}\sum_{i}n_{i\downarrow}n_{i\uparrow},    
\end{equation}
where the Hubbard $U$ is selected via the Feshbach resonance. This is the analogue of the Hubbard $U$ due to Coulomb repulsion in the condensed matter system to be simulated. An approximate form for $U$ is \cite{bloch2008a},
\begin{equation}
U_{\rm Fesh}\approx\sqrt{8}ka_{s}E_{\rm rec}^{1/4}V_{0}^{3/4},
\end{equation}
where,
%
\begin{equation}
a_{s} = a_{s0}\frac{1-\Delta B (B-B_{\mathrm{Res}})}{(B-B_{\mathrm{Res}})^2+\gamma^2/4}
\end{equation}

(we take $a_{s0}=90 \; a_0$, where $a_0$ is the Bohr radius).


\subsection{Full Hamiltonian}

Thus, the full Hamiltonian of the quantum simulator is an analogue of an \emph{extended} HHM,
\begin{equation}
H_{\rm QS} = H_{\rm hop} + H_{\rm R-ph} + H_{\rm ph} + H_{\rm Hub}.
\label{eqn:model}
\end{equation}
The conditions under which this reduces to a site-local HHM will now be determined.


\subsection{Effective interaction}

The form of the effective (retarded) interaction between dressed Rydberg atoms in the (multi)polaron action \cite{kornilovitch2005a},
\begin{equation}
\Phi_{ii'}=\sum_{j,\nu}f_{ij,\nu} f_{i'j,\nu},
\label{eqn:effectiveinteractionphi}
\end{equation}
is highly sensitive to the spot patterns, as shown in Fig. \ref{fig:phi}. We take the limit $r_{c}\ll a$ so that only near-neighbor (NN) and  next-nearest-neighbor (NNN) terms are necessary in Eq.~(\ref{eqn:effectiveinteractionphi}). 

 Repulsive interactions between NNN sites are a feature of the simulator. These arise when $f_{ij}\propto \zetavec_{\nu,j}\cdot\hat{\rvec}_{ij}$ has a negative sign since the direction of $\hat{\rvec}_{ij}$ is opposite to $\zetavec_{\nu,j}$.
Repulsive phonon-mediated interactions might seem surprising, since the electron-phonon interaction is commonly identified as attractive. However, repulsive electron-phonon interaction is predicted in condensed matter systems \cite{alexandrov2002}.

A bipartite lattice with spot arrangements parallel to the lattice vectors (panel (B)) has an effective interaction displaying 4 fold symmetry and a small repulsive interaction on diagonal NNNs. A bipartite checkerboard pattern with no phonon site on alternate squares produces a very similar pattern (not shown).


Crossed patterns lead to  effective interactions with square symmetry and the smallest off-site terms for centrally placed spots (panel (C)). Since the frequency and mass of the oscillators along the two directions are identical and the oscillators are independent, then the full $\Phi$ is the sum of the $\Phi$ for two parallel spot arrangements (panel (F)), rotated by 90$^{o}$ relative to each other \cite{kornilovitch2005a}. The effective interactions are identical for $45^{o}$ and $90^{o}$ orientations of spots. 

\begin{figure}
\includegraphics[width=0.48\textwidth]{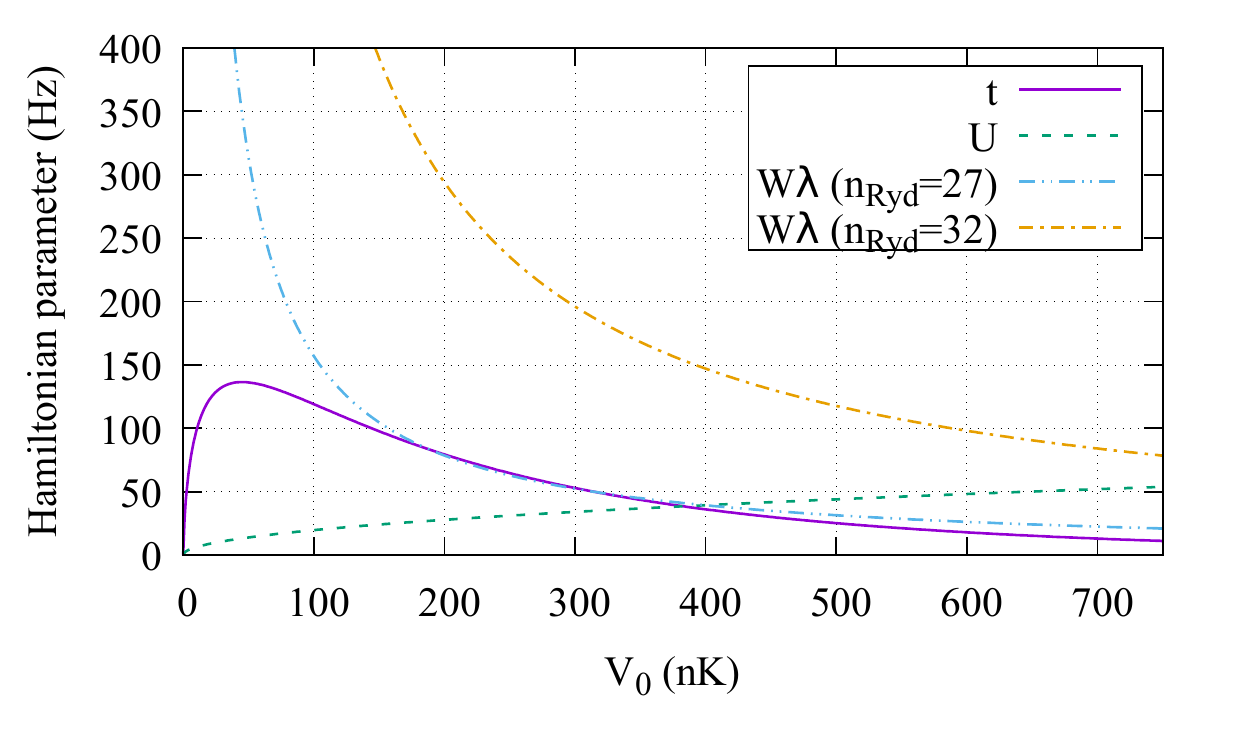}
\caption{[Color online] The intermediate coupling regime with $U, W\lambda \sim t$ can be accessed for $V_{0}\sim 400$ nK, $n_{\rm Ryd}=27$. $W\lambda$ is highly tunable and varies significantly as $n_{\rm Ryd}$ is changed, allowing $W\lambda$ to be changed independently of other interaction parameters. For small $V_{0}$ the hopping dominates the Hubbard $U$ and for large $V_{0}$ the Hubbard $U$ dominates. Thus all orderings of the relative energy scales of interactions can be accessed by varying $V_{0}$ and $n_{\rm Ryd}$.} 
\label{fig:couplings}
\end{figure}

By translating phonon spots so they approach an individual fermion site, $f_{ij}\rightarrow \delta_{ij}$, leading to a better reproduction of the HHM (panels (D)-(F)). As $b$ is decreased from $0.5a'$ (panel (F)) to $0.4a'$ (panel (D)), where $a'=a\sqrt{2}$, the NNN repulsive terms reduce and for $b=0.3a'$, $|\Phi_{\rm NNN}|/|\Phi_{00}| < 5\times 10^{-4}$ (not shown). So reproduction of the HHM, with its local coupling, depends on the lattice spacing and temperature that can be achieved (since larger systems have lower energy scales relative to their condensed matter counterparts \cite{hague2017a}). For comparison, panel (A) shows $\Phi_{ii'}$ for the Holstein model. 

The Rydberg-Rydberg interaction is tunable by selecting different Rydberg states and thus modifying the dipole-dipole interactions. For convenience, we consider states with $n_{\rm Ryd}=n_\mathrm{Rb} = n_\mathrm{K}$. The two-atom state is $|n_\mathrm{Rb} L_\mathrm{Rb};n_\mathrm{K}L_\mathrm{K}\rangle$ where $L_\mathrm{atom}$ is the orbital angular momentum of the atom. Suitable interactions occur for the range $n_{\rm Ryd}\approx 27 - 32$, via the channel $|n_{\rm Ryd}S;n_{\rm Ryd}S\rangle\rightarrow \;  |n_{\rm Ryd}S;(n_{\rm Ryd}-1)P\rangle$. There, $C_6$ ranges from $26.1 \;\mathrm{MHz} \; \mu\mathrm{m}^6$ to $153 \;\mathrm{MHz} \; \mu\mathrm{m}^6$, and the energy difference between $|n_{\rm Ryd}Sn_{\rm Ryd}S\rangle$ and $|n_{\rm Ryd}P(n_{\rm Ryd}-1)P\rangle$ decreases from $6303$ MHz to $4800$ MHz. For $a=1.73$ $\mu$m, a $1/r^6$ potential describes the interactions well.

For convenience, we can define a dimensionless electron-phonon coupling,
\begin{equation}
    \lambda=\frac{\Phi_{00}}{2WM_{\rm Rb}\omega_{\rm ph}^2}.
    \label{eqn:lambdadef}
\end{equation}
$W\lambda$ is a measure of the effective fermion-fermion interaction mediated by phonons and $W=4t$.

 By changing $n_{\rm Ryd}$ and $V_{0}$, the strength of the Rydberg-phonon interaction and ratio $U/t$ can be tuned so that the most interesting regime where $W\lambda\sim U\sim t$ can be explored. In such a regime theoretical techniques often fail, and quantum simulation would be of high value. Fig.  \ref{fig:couplings} shows Hamiltonian parameters for $a=1.73$ $\mu$m, and various $V_{0}$. $t$ and $U$ are of order 100 Hz. For example, for $V_{0}\sim 400$ nK and $n_{\rm Ryd}=27$ the interactions are all of similar strength.

\subsection{Approximate form for NNN repulsion}

For ease of experimental use, we derive an approximate form for the NNN repulsion that is straightforward to calculate without numerically computing $\Psi$ by carrying out the sum.

For the off-center system in Fig. \ref{fig:patterns}(a), the relative size of the interactive term can be estimated if $r_{c}\ll b < a/\sqrt{2}$ and atoms are well localized to sites,
\begin{equation}
\frac{|\Phi_{nnn}|}{|\Phi_{0}|} = \frac{b^{\eta+1}}{(a\sqrt{2}-b)^{\eta+1}}
\end{equation}
i.e. the repulsive term becomes smaller with $b$. To obtain the effective interaction of the HHM, the value of $\eta$ should be as large as possible. For van der Waals terms, $\eta=6$, and thus the decrease in interaction strength with distance is more rapid. Another advantage of the van der Waals term is that they only have weak angular dependence. For $b=a/2\sqrt{2}$, $|\Phi_{nnn}|/|\Phi_{0}| < 5\times 10^{-4}$ and the system closely approximates a pure HHM. 

\begin{figure}
\includegraphics[width=85mm]{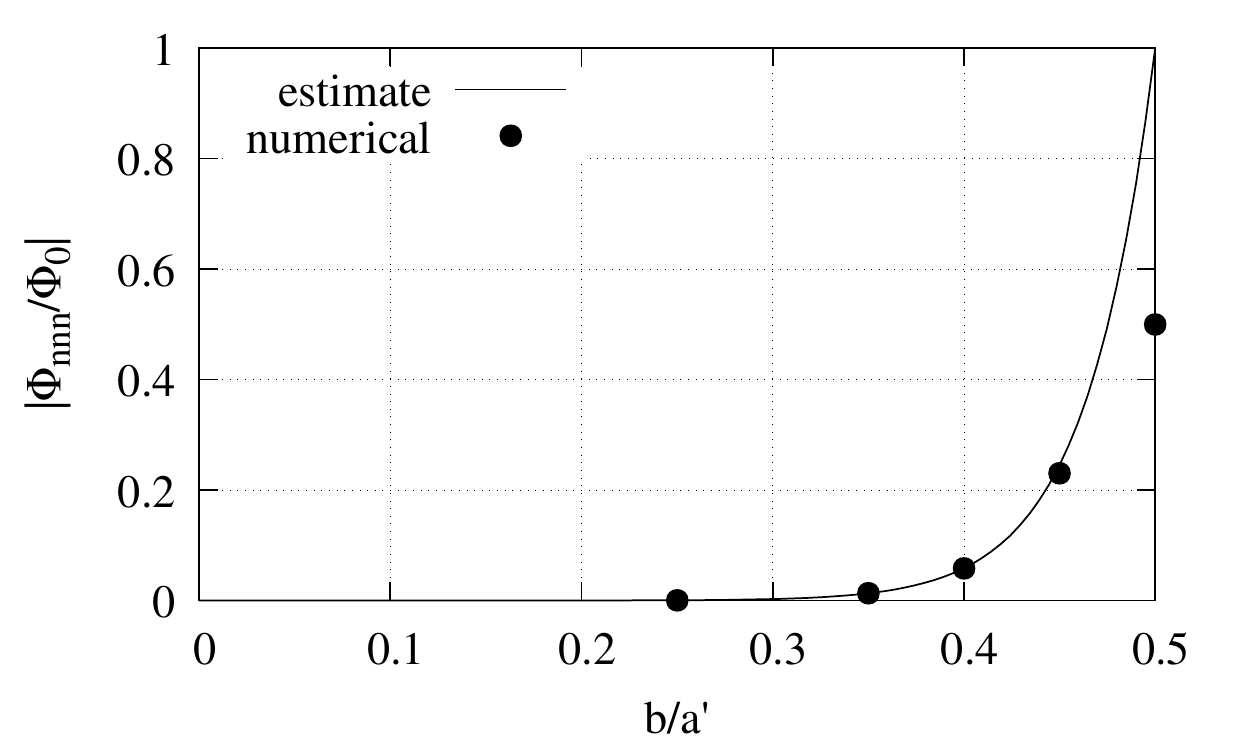}
\caption{Analytical estimates and numerical values of the strength of repulsive NNN relative to onsite terms.}
\label{fig:bovera}
\end{figure}

The approximate form for NNN repulsion can be tested by comparing 
estimated and numerical values of $|\Phi_{nnn}|/|\Phi_{0}|$, Fig. \ref{fig:bovera}. Excellent agreement between numerical and estimated values are found for $b/a'<\sim 0.45$.

\section{Phase diagram}
\label{sec:phase}

To assess the effects of phonon-mediated repulsive interactions on pairing, we examine the limit of large phonon frequency by making a canonical Lang--Firsov transformation \cite{langfirsov}. Application of this transformation to Eq. (\ref{eqn:model}) leads to an effective Hamiltonian,
\begin{eqnarray}
  H_{\rm LF} & = & -t'\sum_{\langle ij\rangle}c^{\dagger}_{i}c_{j}+\sum_{ii'} n_{i}n_{i'} \frac{W\lambda\Phi_{i,i'}}{\Phi_{00}} + U_{\rm Fesh}\sum_{i}n_{i\downarrow}n_{i\uparrow} \nonumber\\
& &  \hspace{30mm} + \sum_{j\nu} \hbar\omega_{\rm ph,\nu} d^{\dagger}_{j\nu}d_{j\nu},
  \end{eqnarray}
%
When $\hbar\omega_{\rm ph}\gg t$ the effective hopping, $t' = t\exp\left[-W\lambda(1-\Phi_{\rm NN}/\Phi_{00})/\hbar\omega_{\rm ph}\right]$. 

If repulsive interactions are found on both diagonals (Fig. \ref{fig:phi} (B) and (C)), solving the two-body Schr\"odinger equation for $H_{\rm LF}$ establishes the critical coupling, 
\begin{equation} 
U^{(C)}_{\rm Fesh} = - \frac{
\gamma_3 t' V_1 V_2 + 4 t^{\prime 2} ( V_1 + V_2 ) }
                    { \gamma_1 V_1 V_2 + \frac{1}{2} \, t' V_1 + \gamma_2 t' V_2 + t^{\prime 2} } + 2W\lambda \: .  
\end{equation}
where, $\gamma_1 = (32 - 9\pi)/12 \pi, \gamma_2 = (16 - 3\pi)/3\pi, \gamma_3 = (64 - 18 \pi)/3 \pi$, NN
interaction is $V_{1}=-2\Phi_{\rm NN}W\lambda$ and diagonal interaction
$V_{2}=-2\Phi_{\rm NNN}W\lambda$ (see Appendix \ref{app:uvmodelsolve}).

If repulsion is found on a single diagonal (Fig. \ref{fig:phi}(D-F)),
\begin{equation}
U^{(C)}_{\rm Fesh}=\frac{4W\lambda\Phi_{\rm NNN}t'}{t'-8W\lambda\Phi_{\rm NNN}/3\pi} + 2W\lambda .
\end{equation}

\begin{figure}
\includegraphics[width=0.48\textwidth]{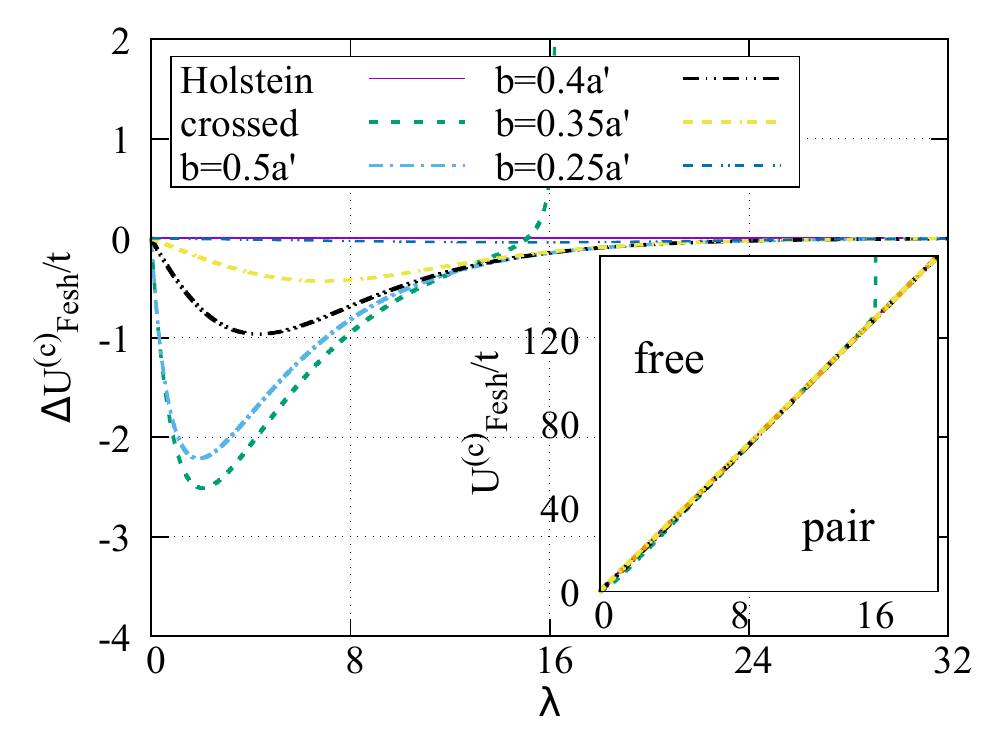}
  \caption{[Color online] Repulsive terms lead to small differences between $U^{(C)}$ in the HHM and quantum simulator, $\Delta U^{(C)}$, which are already $\lesssim 4\%$ for $b = 0.4a'$. The divergence for the crossed spot configuration at $\lambda\sim 16.35$ is due to stabilizing effects of attractive NN coupling. 
    }
  \label{fig:threshold}
  \end{figure}

\begin{figure}
\includegraphics[width=0.48\textwidth]{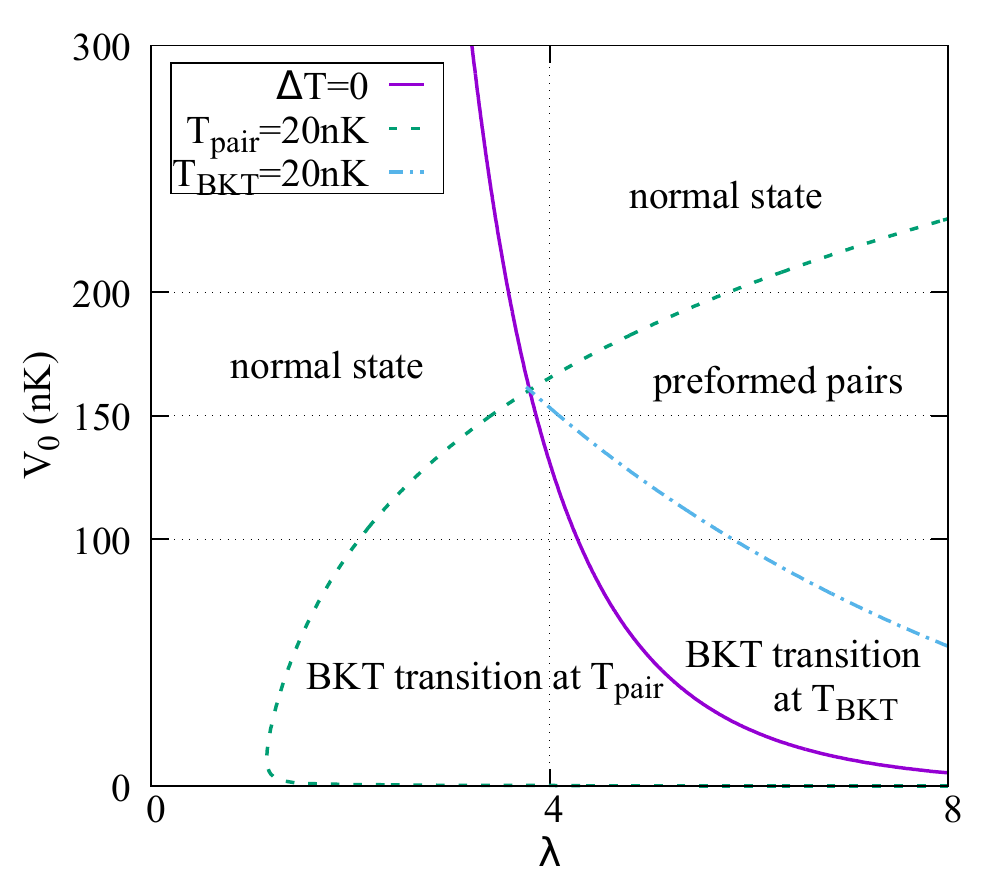}
  \caption{[Color online] At 20nK within the strong coupling theory a range of paired, normal and BKT states are accessible to experiment. The line $\Delta T=T_{\rm BKT}-T_{\rm pair}=0$ shows where temperatures for pairing and BKT transition are equal. $D=0.2823\mu$m, $V_{0,\rm ph}=2.5 V_{0}$ so $\hbar\omega_{\rm ph} = 18.52t$.}
  \label{fig:pairingtemperatures}
  \end{figure}

At intermediate $\lambda$, the binding diagram is essentially unchanged from the HHM. In Fig. \ref{fig:threshold}, $U^{(C)}$ is calculated for the $\Phi$ shown in Fig. \ref{fig:phi}. Attractive NN terms push $U^{(C)}\rightarrow\infty$ at finite $\lambda$. Repulsive terms decrease $U^{(C)}$.





The phase diagram at 20 nK shown in Fig.~\ref{fig:pairingtemperatures} has four distinct regions. If $T_{\rm pair}<T_{\rm BKT}$ there is condensation at the BKT temperature. At $T_{\rm pair}>T_{\rm BKT}$, preformed pairs condense at $T_{\rm BKT}$. We predict a region of the parameter space with preformed pairs for $T_{\rm BKT}< T < T_{\rm pair}$. The normal state is at $T > T_{\rm pair}$.

We predict that in experiments, phonon-mediated local pairing occurs at $\sim 20$ nK for $\lambda\gtrsim 1.5$. Local pairs can be directly observed using gas microscopy \cite{mitra2018a}. For $\lambda\gg 1$, local s pairs dominate below $T_{\rm pair}\sim (2W\lambda-8t')/k_{B}$. Figure~\ref{fig:pairingtemperatures} shows how local pairing occurs at $\sim 20$ nK for large $\lambda$, small $V_{0}$. These $\lambda$ values are large compared to condensed matter analogues, but are accessible in the quantum simulator at large $n_{\rm Ryd}$.

Using an expression for the effective mass, calculated at large $\lambda$, we predict that BKT temperatures of $\sim 20$ nK can be achieved in experiments at small $V_{0}$ (Fig.~\ref{fig:pairingtemperatures}). BKT condensation would be identifiable via changes to the momentum distribution of the atoms, which can be measured using time of flight.  No general expression exists for the BKT temperature, so we make estimates for low pair density $n_{B}=0.01\ll 1$ where $T_{{\rm BKT}}=4\pi\hbar^2 n_{B}/a^2 k_{B}2m^{**}\ln\ln(4/n_{B})$ \cite{fisher1988,alexandrov2013}.
%
For strongly coupled onsite pairs (large $\lambda$), effective pair mass
%
$m^{**}=\hbar^2\sqrt{W^2\lambda^2+2t'^{2}}/t^{\prime 2} a^2$ (see Appendix \ref{app:mass}). 

We, therefore, predict that it is possible to transition between normal, preformed pair, and BKT phases at $\sim 20$ nK by selecting $V_{0}=150$ nK, $n_{\rm Ryd}=34$ and $\bar{\alpha}<0.09961$ to get $\lambda\lesssim 5$. For lower temperatures, this can be done at smaller $\lambda$. Thus, the proposed simulator offers a route to the (as yet) unexplored physics of boson-mediated pairing and condensation of fermions in cold atom quantum simulators.

\section{Conclusions}
\label{sec:conclusions}

We have proposed a quantum simulator for boson mediated pairing, and demonstrated that preformed pairs and a BKT transition are expected in certain limits of the parameter space of the simulator at a temperature of 20 nK. We predict that it is possible to carry out quantum simulation of the transition between normal, preformed pair, and BKT phases at $\sim 20$ nK in the experiment that we have proposed. 

The proposed quantum simulator has potential to provide insight regarding the origins of the pseudogap in unconventional superconductors. The leading hypotheses are that: (a) that the pseudogap appears at the same temperature as preformed pairs (b) that the pseudogap occurs due to fluctuations unrelated to the superconductivity (e.g. spin fluctuations, charge density waves) (c) a hybrid of both views with two gaps (see e.g. \cite{vishik2018}). Evidence for and against all of these viewpoints can be found using differing experimental techniques. Since our prediction is that preformed pairs are accessible within the quantum simulator at large $\lambda$, then it may be possible to use the quantum simulator to probe the extent to which preformed pairs are consistent with a pseudogap in a controlled, tunable manner.

Our calculations for the phase diagram are valid for large coupling and phonon frequencies, so experiments are needed to examine the phases for more modest $\lambda\sim 1$, $\hbar\omega_{0}\sim t$ (as e.g. found in cuprate superconductors). The trend of the phase diagrams at strong coupling indicates that the preformed pair state is likely to be found for smaller $\lambda$ at lower temperatures. 

While quantum simulators have been constructed to investigate other questions in superconductivity, such as the BCS-BEC crossover, these are purely fermionic in nature, and as such the interactions between the fermions (which are mediated by the Feshbach resonance) are instantaneous. The proposed simulator would permit the investigation of BCS-BEC crossover in a distinct and more realistic regime where the interactions between the fermions are retarded. The quantum simulator would also enable other problems regarding retardation to be investigated, such as the adiabatic polaron.

So, in conclusion, we expect that the proposed simulator offers a route to the (as yet) unexplored physics of boson-mediated pairing and condensation of fermions in cold atom quantum simulators, and may have the capability to explore the possible relationship between pseudogaps and preformed pairs in unconventional superconductors.



\bibliographystyle{apsrev4-1}
\bibliography{cupraterefs}

\appendix
\section{Solution of $UV$ model}
\label{app:uvmodelsolve}

In this appendix, we solve $UV$ models for two cases pertinent to the current quantum simulator: (1) The case where there is $V$ only on a single diagonal and (2) the case where there is $V$ on both diagonals.


%
%

\subsection{\label{QSM:sec:two}
$V$ on a single diagonal
}
\label{app:uv}

\begin{figure}
\includegraphics[width=0.40\textwidth]{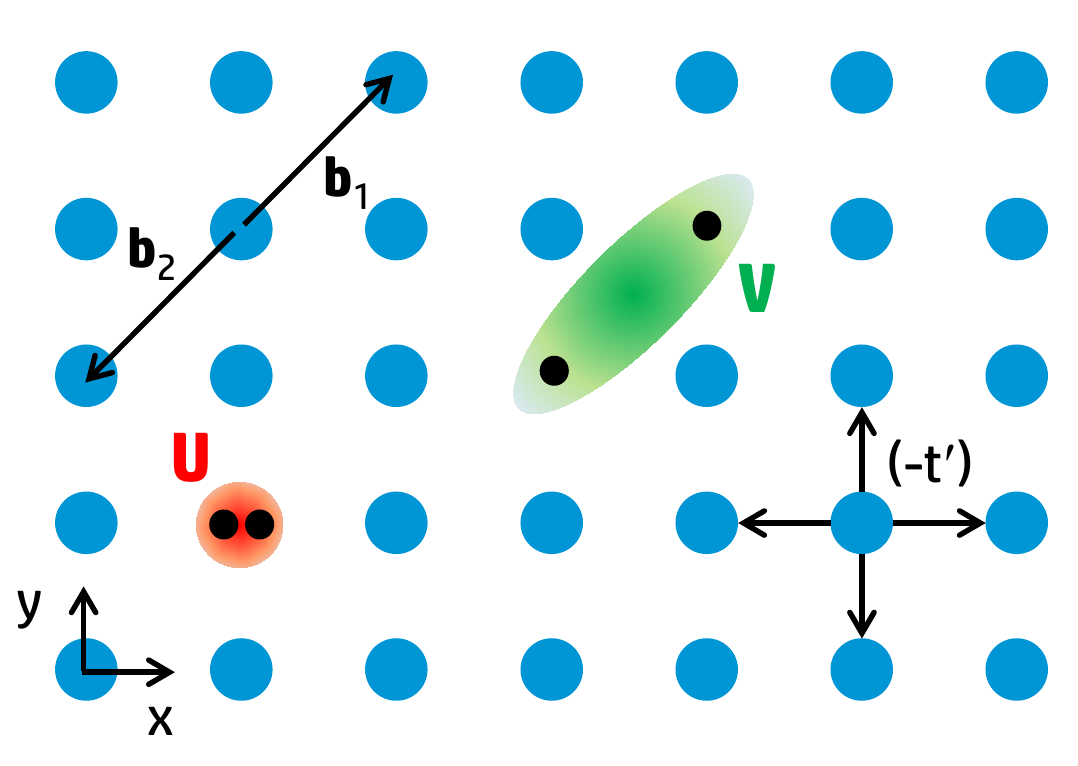}
\caption{[Color online] Schematic of $UV$ model with $V$ on a single NNN diagonal only and no interaction on NN, consistent with Fig. \ref{fig:phi}, panels (D-F).}
\label{QSM:fig:one}
\end{figure}

In the ``diagonal'' model two atoms interact with potential $V$ if they are separated by NNN vectors ${\bf b}_1 = + ( {\bf x} + {\bf y})$ or ${\bf b}_2 = - ( {\bf x} + {\bf y})$, as shown schematically in Fig.~\ref{QSM:fig:one}. The Hamiltonian is,
\begin{eqnarray}
H_{\rm diag}  & = & - t' \sum_{\langle {\bf m m'} \rangle, \sigma} 
             c^{\dagger}_{{\bf m} \sigma} c_{{\bf m'}, \sigma}   
+ \frac{U}{2} \sum_{\bf m} \hat{n}_{\bf m} \left( \hat{n}_{\bf m} - 1 \right)              
\nonumber \\    
    &  & + \frac{V}{2} \sum_{\bf m} \sum_{{\bf b} = {\bf b}_{1,2}} 
    \hat{n}_{\bf m} \hat{n}_{{\bf m} + {\bf b}} \: .     
\label{QSM:eq:three}    
\end{eqnarray}
Here, {\bf m} indexes lattice sites, $\langle {\bf m m'} \rangle$ are pairs of NNs, $\sigma = \pm \frac{1}{2}$ is the $z$-axis spin projection, $\hat{n}_{\bf m} = \sum_{\sigma} c^{\dagger}_{{\bf m} \sigma} c_{{\bf m} \sigma}$ is the total fermion number operator on site ${\bf m}$, and ${\bf b}$ are lattice vectors with nonzero interaction between atoms. The atom's kinetic energy is defined by a dispersion law
\begin{equation}
\varepsilon_{\bf k} = - 2t' \left( \cos{k_x} + \cos{k_y} \right) .  
\label{QSM:eq:two}
\end{equation}

%

\begin{figure}
\includegraphics[width=0.48\textwidth]{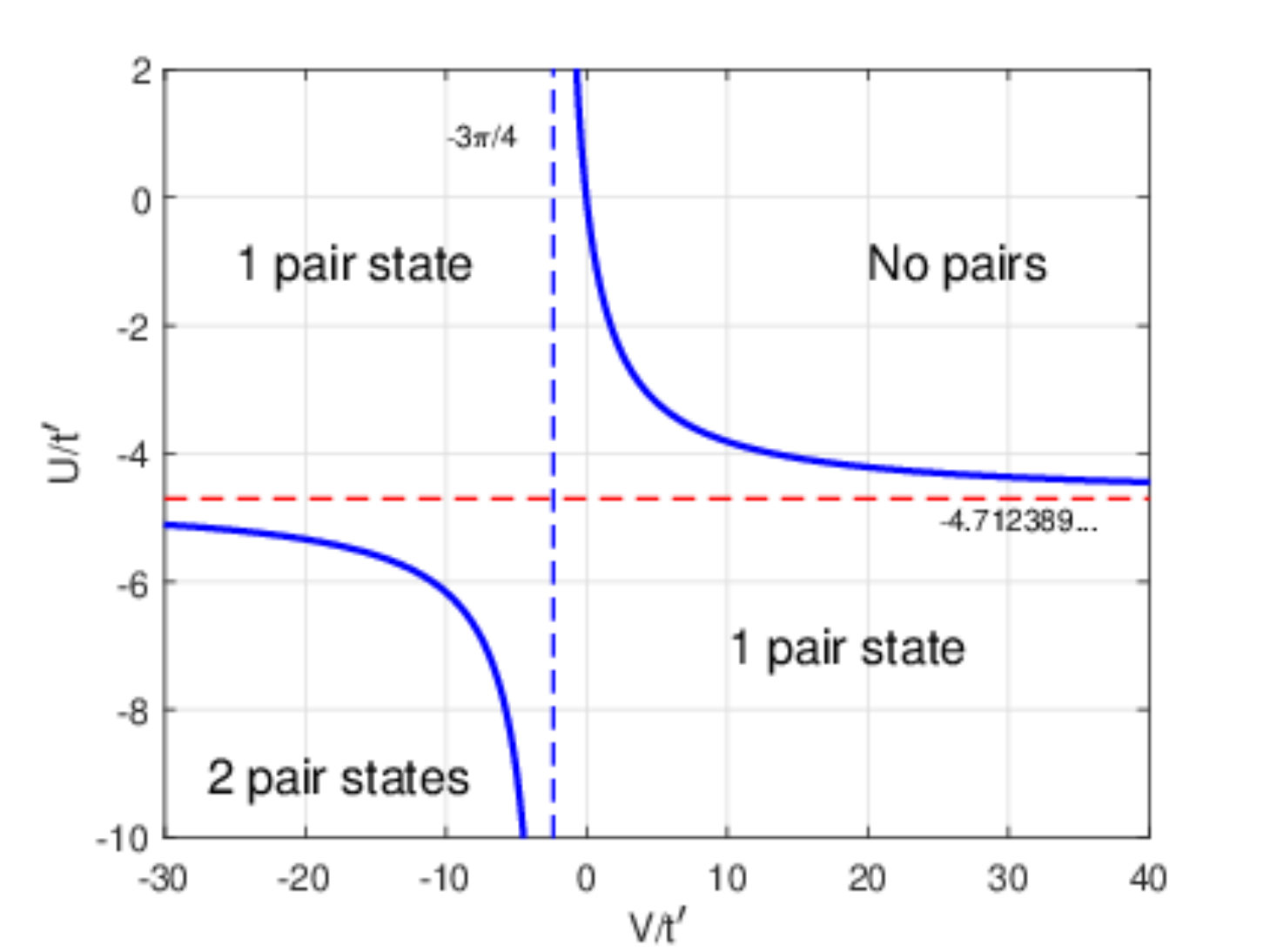}
\caption{[color online] Phase diagram of model (\ref{QSM:eq:three}). Dashed lines show asymptotes.}
\label{QSM:fig:two}
\end{figure}

The two-body case of model (\ref{QSM:eq:three}) is now solved for symmetrical wave function $\Psi({\bf r}_1, {\bf r}_2) = \Psi({\bf r}_2, {\bf r}_1)$ with zero total momentum ${\bf K} = {\bf k}_1 + {\bf k}_2 = 0$. The method is described in detail elsewhere~\cite{Kornilovitch2019}. The pair energy $E$ is found from the $(2 \times 2)$ determinant equation:    
\begin{equation}
\left\vert \begin{array}{cc} 
U M_{00} + 1  & 2 V M_{11} \\
U M_{11}      & V ( M_{00} + M_{22} ) + 1 
\end{array} \right\vert = 0 \: ,  
\label{QSM:eq:four}
\end{equation}
where
\begin{equation}
M_{nl} = \int\limits^{\pi}_{-\pi} \!\! \int\limits^{\pi}_{-\pi} \frac{dq_x \, dq_y}{(2\pi)^2} 
\frac{\cos{nq_x} \cos{lq_y}}{ \vert E \vert - 4t' ( \cos{q_x} + \cos{q_y} ) }  \: .  
\label{QSM:eq:five}
\end{equation}
All $M_{nm}$ can be expressed via complete elliptic integrals of the first kind $K(\kappa)$ and second kind $E(\kappa)$ utilizing the two types of recurrence relations \cite{Morita1975,Joyce2002}. Relevant results are 
\begin{align*}
M_{00} & = \frac{2}{\pi |E|} K( \kappa ) \: ,
  \\
M_{10} & = \frac{1}{\pi W'} K( \kappa ) - \frac{1}{2W'} \: ,
  \\
M_{11} & = \frac{|E|}{2 \pi W^{\prime 2}} \left\{ ( 2 - \kappa^2 ) K( \kappa ) - 2 E( \kappa ) \right\} \: ,
 \\
M_{20} & = \frac{2}{\pi |E|} K( \kappa ) +
\frac{|E|}{W^{\prime 2}} \left\{ \frac{2}{\pi} \, E( \kappa ) - 1 \right\} \: ,
  \\
M_{21} & = \left( \frac{|E|^2}{\pi W^{\prime 3}} - \frac{3}{\pi W'} \right) K( \kappa )
- \frac{|E|^2}{\pi W^{\prime 3}} E( \kappa ) + \frac{1}{2W'} \: ,
 \\
M_{22} & = \left( \frac{2}{\pi |E|} - \frac{8 |E|}{3 \pi W^{\prime 2}} + 
                    \frac{ 2 |E|^3 }{ 3 \pi W^{\prime 4} } \right) K( \kappa ) + 
\nonumber \\
       &  \quad + \left( \frac{4 |E|}{3 \pi W^{\prime 2}} - \frac{2 |E|^3}{3 \pi W^{\prime 4}} \right) E( \kappa ) \: ,
\label{QSM:eq:fivesix}
\end{align*}
where $\kappa \equiv 2W'/|E| \leq 1$ and  $W' = 4t'$. It is also convenient to introduce differences:
\begin{equation}
C_{nl} = \int\limits^{\pi}_{-\pi} \!\! \int\limits^{\pi}_{-\pi} \frac{dq_x \, dq_y}{(2\pi)^2} 
\frac{1 - \cos{nq_x} \cos{lq_y}}{ \vert E \vert - 4t' ( \cos{q_x} + \cos{q_y} ) }  \: ,  
\label{QSM:eq:six}
\end{equation}
so that 
\begin{equation}
M_{nl} = M_{00} - C_{nl} \: .  
\label{QSM:eq:seven}
\end{equation}
Substituting Eq.~(\ref{QSM:eq:seven}) in Eq.~(\ref{QSM:eq:four}) and expanding the determinant yields the dispersion equation:
\begin{eqnarray}
M_{00} \left[ U + 2V + UV ( 4C_{11} - C_{22} ) \right]  &   &
\nonumber \\   
     + \left[ 1 - V C_{22} - 2UV \, C^2_{11} \right]    & = & 0 \: .      
\label{QSM:eq:eight}    
\end{eqnarray}
To find the pairing threshold, we set $E \rightarrow -8t' - 0$. Then the base integral $M_{00}$ diverges logarithmically, so that Eq.~(\ref{QSM:eq:eight}) reduces to
\begin{equation}
U + 2V + UV ( 4C_{11} - C_{22} ) = 0 \: .  
\label{QSM:eq:nine}
\end{equation}
In the same limit, integral differences $C_{11}$ and $C_{22}$ converge. Elementary integration yields  
\begin{align}
      C_{11}(-8t') & = \frac{1}{2\pi t'} \: ,
\label{QSM:eq:ten} \\   
      C_{22}(-8t') & = \frac{2}{3\pi t'} \: ,
\label{QSM:eq:eleven} \\   
4 C_{11} - C_{22} & = \frac{4}{3\pi t'} \: .      
\label{QSM:eq:twelve}    
\end{align}
The binding condition takes the final form
\begin{equation}
U + 2V + \frac{4}{3\pi} \frac{UV}{t'} = 0 \: .  
\label{QSM:eq:thirteen}
\end{equation}
If both $U$ and $V$ are positive, Eq.~(\ref{QSM:eq:thirteen}) does not have a solution: all states are non-bound. If either $U$ or $V$ is negative, there is a bound state if the attractive potential exceeds a threshold. For example, if $V > 0$ the pair is formed if 
\begin{equation}
U < U_{\rm cr} = - \frac{2Vt'}{ t' + \frac{4}{3\pi} V } \: .  
\label{QSM:eq:fourteen}
\end{equation}
The function $U_{\rm cr}(V)$ is shown in figure~\ref{QSM:fig:two}. In the limit $V \rightarrow \infty$, the binding threshold is $U_{\rm cr} \rightarrow - (6\pi/4) \, t' = (- 4.712389 \ldots) \, t'$.

\begin{figure*}
\includegraphics[width=0.48\textwidth]{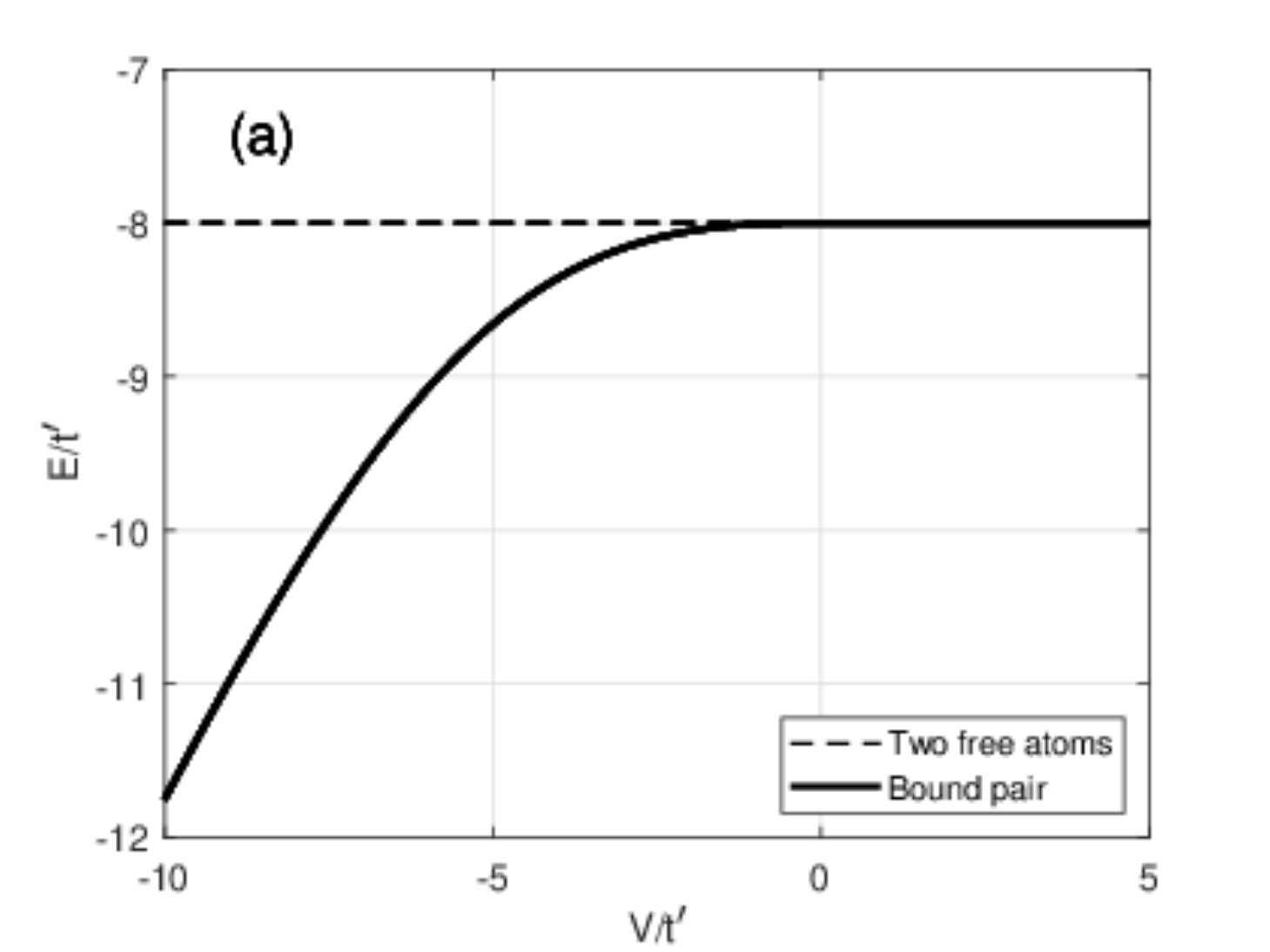}
\includegraphics[width=0.48\textwidth]{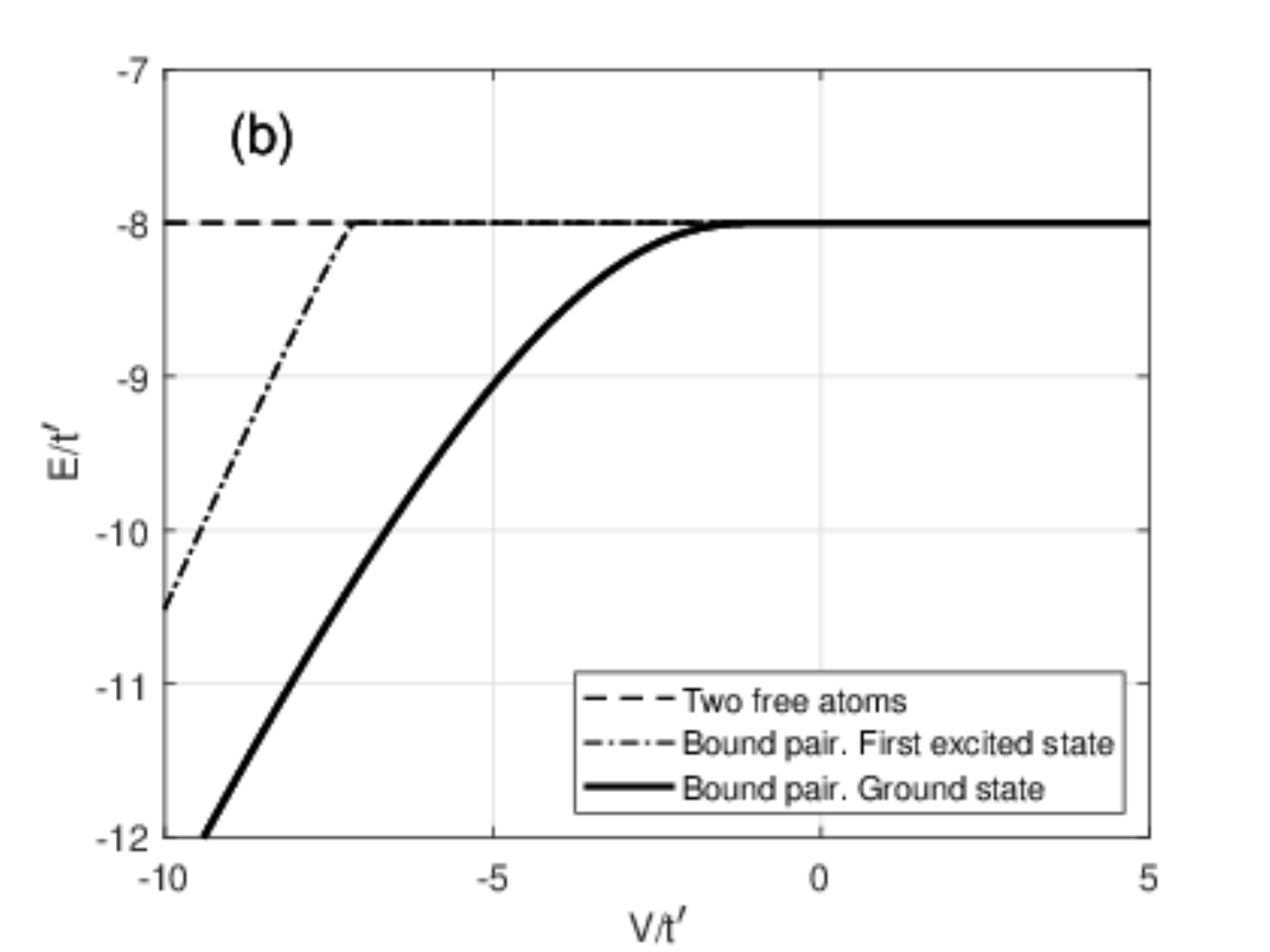}
\caption{Pair energy when $V$ is only on a single diagonal (\ref{QSM:eq:three}) for (a) $U = -2t'$. (b) $U = V$.}
\label{QSM:fig:twoone}
\label{QSM:fig:twotwo}
\end{figure*}

If $U$ and $V$ are both negative, a second bound state may appear. The corresponding threshold can also be deduced from Eq.~(\ref{QSM:eq:thirteen}).  

The existence of either one or two pair states can be validated by directly solving the dispersion equation~(\ref{QSM:eq:four}). Figure~\ref{QSM:fig:twoone}(a) shows the pair dispersion as a function of $V$ for a fixed value $U = -2t'$. There is only one state, which agrees with the phase diagram of Fig.~\ref{QSM:fig:two}. However, along the line $U = V$ the phase diagram predicts the existence of a second state for $V < -7.1 \, t'$. A corresponding pair dispersion is shown in Fig.~\ref{QSM:fig:twotwo}(b), which clearly shows two pair branches.

\subsection{\label{QSM:sec:three}
$V$ on both diagonals
}

\begin{figure}
\includegraphics[width=0.4\textwidth]{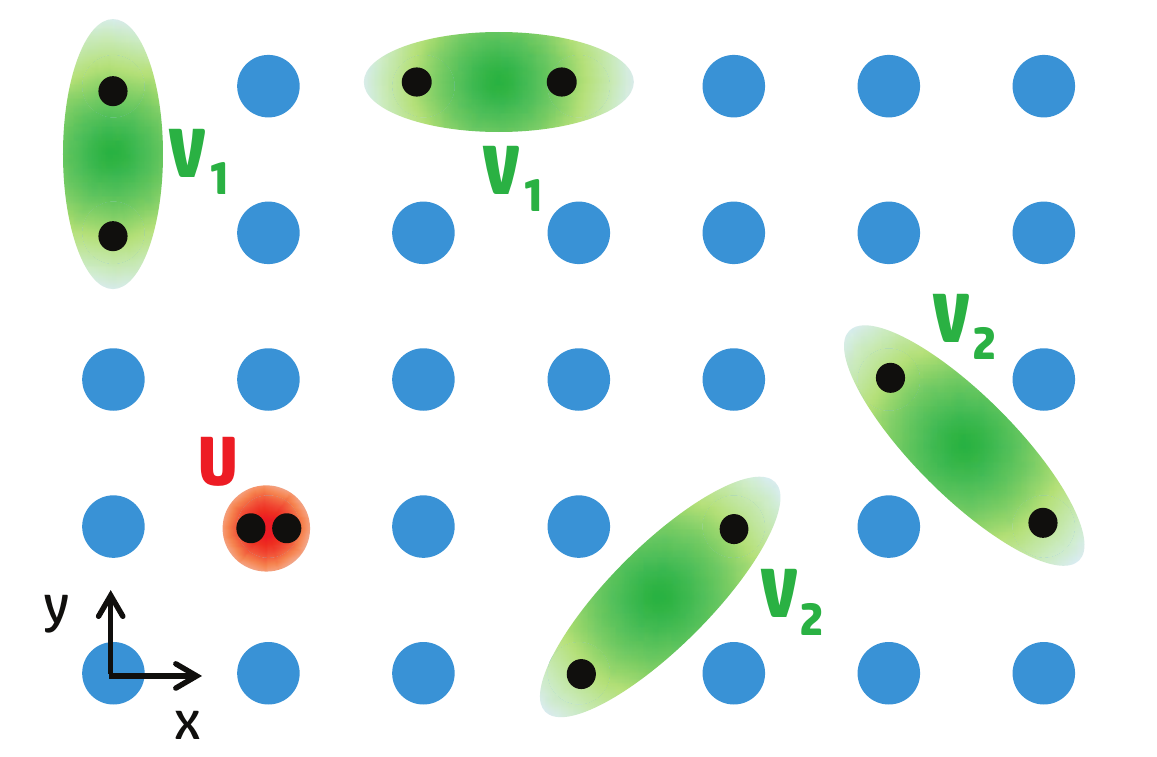}
\caption{[Color online] Schematic of $UV$ model with NNN $V$, and no NN interaction, consistent with Fig. \ref{fig:phi}(B) and (C). }
\label{QSM:fig:three}
\end{figure}

 In this version of the model, the atoms interact with potential $U$ if occupy the same site, with potential $V_1$ if separated by one of {\em four} NN vectors ${\bf b}_1 = \pm {\bf x}$ or $\pm {\bf y}$, and with potential $V_2$ if separated by one of {\em four} next-nearest vectors ${\bf b}_2 = \pm ( {\bf x} \pm {\bf y} )$. A schematic of the model is shown in Fig.~\ref{QSM:fig:three}. The ground state energy is determined from the $( 3 \times 3 )$ determinant equation   

\begin{widetext}
\begin{equation}
\left\vert \begin{array}{ccc} 
U M_{00} + 1  & 2 V_1 M_{10}                             &  2 V_2 M_{11}                \\
U M_{11}      & V_1 ( M_{00} + M_{20} + 2 M_{11} ) + 1   &  2 V_2 ( M_{10} + M_{21} )   \\
2 U M_{11}    & 2 V_1 ( M_{10} + M_{21} ) &  V_2 ( M_{00} + M_{22} + 2 M_{20} ) + 1 
\end{array} \right\vert = 0 \: .  
\label{QSM:eq:fifteen}
\end{equation}
\end{widetext}
where $M_{nl}$ are defined in Eq.~(\ref{QSM:eq:five}). Introducing differences (\ref{QSM:eq:six}) and expanding the determinant, one obtains, similarly to Eq.~(\ref{QSM:eq:eight}), $A \cdot M_{00} A + B = 0$, where $A$ and $B$ are complicated expressions. ($A$ is given below.) To obtain the threshold, set $E \rightarrow -8t' - 0$ where $M_{00}$ diverges. Thus, the binding condition reduces to $A = 0$, or in full form  
\begin{align}
A = & ( U + 4 V_1 + 4 V_2 ) \nonumber\\ 
&+  U V_1 ( 8 C_{10} - 2 C_{11} - C_{20} ) \nonumber\\ 
& +  U V_2 ( 8 C_{11} - 2 C_{20} - C_{22} ) + \nonumber\\
& +  V_1 V_2 ( 8 C_{11} + 12 C_{20} - 16 C_{21} + 4 C_{22}  -16 C_{10})\nonumber\\
& +  U V_{1} V_{2} (2 C^2_{20} - 4C^2_{10} - 32 C^2_{11} - 4 C^2_{21} + 48 C_{10} C_{11}\nonumber\\
& \qquad- 16 C_{10} C_{20} + 8 C_{10} C_{21} - 4 C_{11} C_{20}- 8 C_{10} C_{22}\nonumber \\
& \qquad + 16 C_{11} C_{21} + 2 C_{11} C_{22} + C_{20} C_{22} ) = 0 \: . 
\label{QSM:eq:sixteen}    
\end{align}
In addition to $C_{11}$ and $C_{22}$ given in Eqs.~(\ref{QSM:eq:ten}) and (\ref{QSM:eq:eleven}), one needs the following integrals:
\begin{eqnarray}
C_{10}(-8t') & = & \frac{1}{8 t'} \: ,
\label{QSM:eq:seventeen} \\   
C_{20}(-8t') & = & \frac{\pi - 2}{2\pi t'} \: ,
\label{QSM:eq:eighteen} \\   
C_{21}(-8t') & = & \frac{8 - \pi}{8\pi t'} \: .      
\label{QSM:eq:nineteen}    
\end{eqnarray}
Substituting everything in Eq.~(\ref{QSM:eq:sixteen}) one obtains the binding condition
\begin{eqnarray}
\gamma_1 U V_1 V_2 + \frac{1}{2} \, t' U V_1 +& &\nonumber
\gamma_2 t' U V_2 + \gamma_3 t' V_1 V_2 + \\
t^{\prime 2} ( U + 4 V_1 + 4 V_2 ) & =  &0 \: ,  
\label{QSM:eq:twenty}
\end{eqnarray}
where 
\begin{eqnarray}
\gamma_1 & = & \frac{ 32 - 9\pi }{ 12 \pi }  = 0.0988263632 \ldots \: ,
\label{QSM:eq:twentyone} \\   
\gamma_2 & = & \frac{ 16 - 3\pi }{ 3\pi }    = 0.6976527263 \ldots \: ,
\label{QSM:eq:twentytwo} \\   
\gamma_3 & = & \frac{ 64 - 18 \pi }{ 3 \pi } = 0.7906109053 \ldots \: .      
\label{QSM:eq:twentythree}    
\end{eqnarray}
The binding condition (\ref{QSM:eq:twenty}) does not have a solution when $U$, $V_1$, $V_2$ are all positive. Equation~(\ref{QSM:eq:twenty}) only has a nontrivial solution if the interaction potentials are of different signs. For example, if $V_1, V_2 > 0$, then the critical value of $U$ is strictly negative
\begin{equation}
U_{\rm cr} = - \frac{ \gamma_3 t' V_1 V_2 + 4 t^{\prime 2} ( V_1 + V_2 ) }
                    { \gamma_1 V_1 V_2 + \frac{1}{2} \, t' V_1 + \gamma_2 t' V_2 + t^{\prime 2} } \: .  
\label{QSM:eq:twentyfour}
\end{equation}
Atoms are bound into pairs if $U < U_{\rm cr}$ and unbound otherwise. 

\begin{figure}
\includegraphics[width=0.48\textwidth]{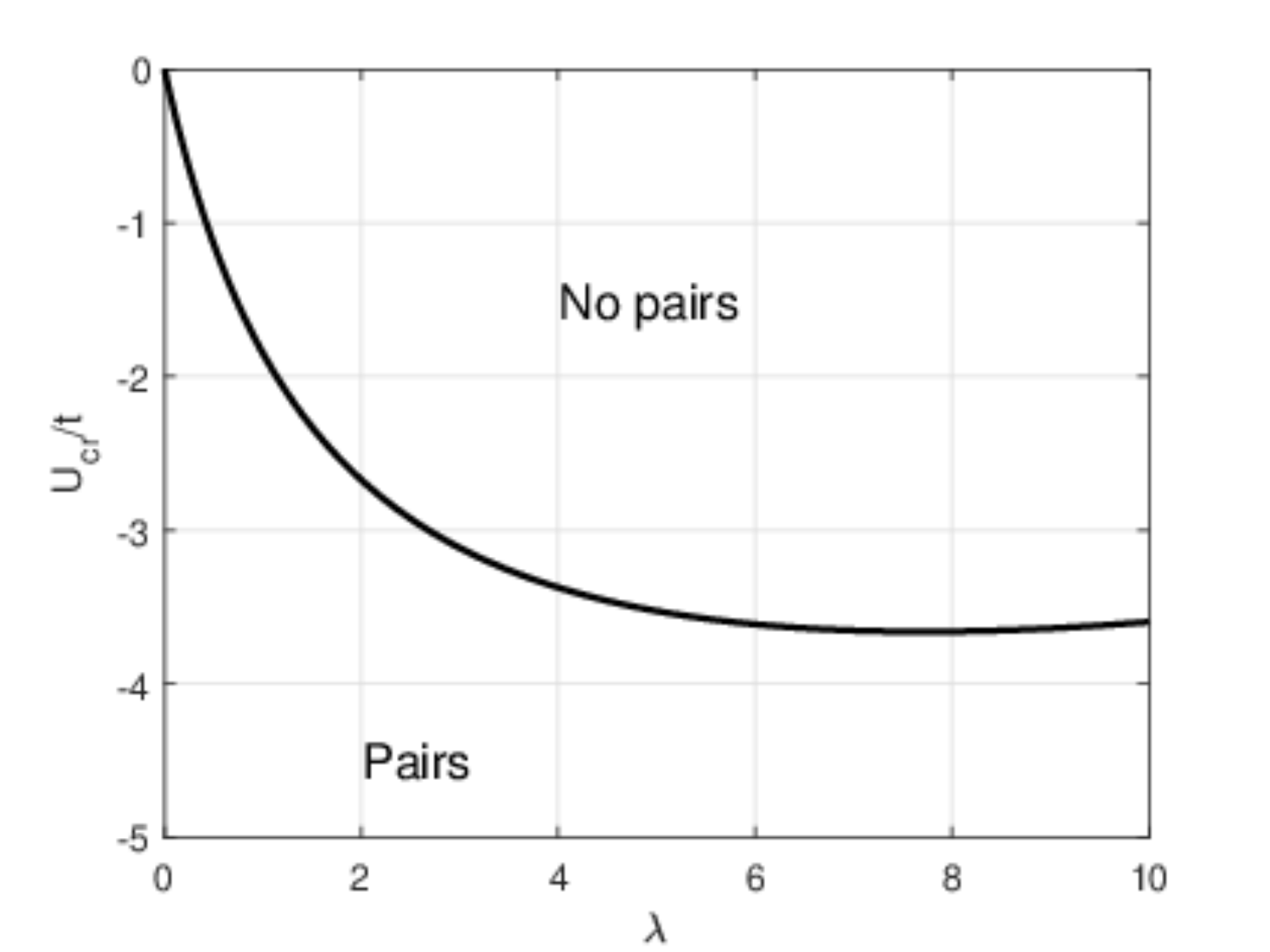}
\caption{The boundary line $U(\lambda)$ derived from Eq.~(\ref{QSM:eq:twentyfive}) for $t' = t$.}
\label{QSM:fig:four}
\end{figure}
\begin{figure}
\includegraphics[width=0.48\textwidth]{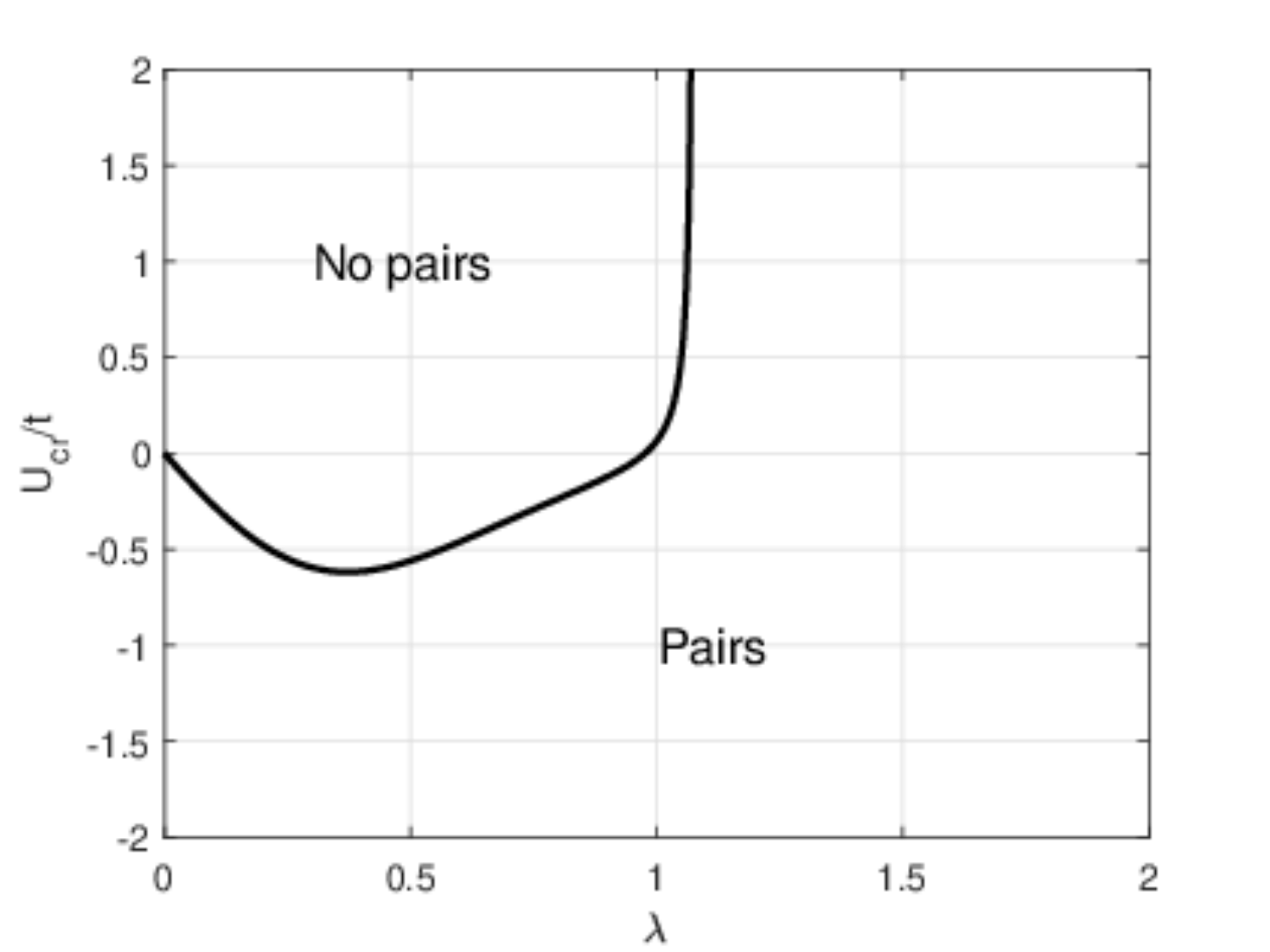}
\caption{The boundary line $U(\lambda)$ derived from Eq.~(\ref{QSM:eq:twentyfive}) but with renormalized hopping integral, Eq.~(\ref{QSM:eq:twentysix}). Only the ground state formation line is shown.}
\label{QSM:fig:five}
\end{figure}

When the Lang--Firsov transformation is applied to derive a $UV$ model from the extended Hubbard--Holstein model, the hopping becomes normalized:
\begin{equation}
t' = t \, e^{ - 4 ( 1 - 0.16 ) \lambda} \: ,   
\label{QSM:eq:twentysix}
\end{equation}
where $t'$ is the renormalized hopping and $t$ the bare (non-renormalized by phonon interaction) hopping integral in the Hubbard--Holstein model.

In a physically relevant case, $V_1 = -0.16 \lambda t$ and $V_2 = 0.896 \lambda t$, where $\lambda$ is a dimensionless coupling constant. In this particular case, Eq.~(\ref{QSM:eq:twentyfour}) takes the form
\begin{equation}
U_{\rm cr} = \frac{       0.1133 \, t^2 t' \lambda^2 - 2.9440 \, t t^{\prime 2} \lambda   }
                  { t^{\prime 2} + 0.5451 \, t   t' \lambda   - 0.0142 \, t^2   \lambda^2 } \: .  
\label{QSM:eq:twentyfive}
\end{equation}
In the case of non-renormalized hopping, $t' = t$, the boundary line $U(\lambda)$ derived from the last expression is shown in Fig.~\ref{QSM:fig:four}. 

If, in addition, the hopping integral is renormalized 
the same dependence changes shape to what is shown in Fig.~\ref{QSM:fig:five}. Note the presence of a singularity near $\lambda = 1.08$. At even larger $\lambda$, a second bound state might appear. The corresponding threshold line is not shown.   
  
Finally, the on-site potential is a sum of Feshbach interaction and phonon-mediated attraction:
\begin{equation}
U = U_{\rm Fesh} - 8 t \lambda \: .   
\label{QSM:eq:twentyseven}
\end{equation}
The pairing line $U_{\rm Fesh}(\lambda)$ is shown in the main body of the paper. 
%
%

\section{Pair mass in $UV$ model at strong coupling}
\label{app:mass}

Application of the Lang--Firsov transformation leads to an effective instantaneous interaction for Hamiltonian \ref{eqn:model},
\begin{equation}
    \tilde{H} = -t'\sum_{ij}c^{\dagger}_{i}c_{j} + \sum_{ij} V_{ij} n_i n_j
\end{equation}
where $V_{ii}=U$ and $V_{i,i+1}=V$. For attractive $U$ and $V$, a trial strong coupling wavefunction is 
\begin{equation}
    |\Psi\rangle = \frac{1}{N}\sum_{i}e^{i\kvec\cdot\rvec_{i}}\left(aA^{\dagger}_{i}+bB^{\dagger}_{i}+cC^{\dagger}_{i}\right)\vacu
\end{equation}
where
\begin{eqnarray}
    A^{\dagger}&=&c^{\dagger}_{\rvec_i\uparrow}c^{\dagger}_{\rvec_i\downarrow}\nonumber\\
        B^{\dagger}&=&c^{\dagger}_{\rvec_i\uparrow}c^{\dagger}_{\rvec_i+\tauvec_{1}\downarrow}\nonumber\\
            C^{\dagger}&=&c^{\dagger}_{\rvec_{i}\uparrow}c^{\dagger}_{\rvec_i+\tauvec_{2}\downarrow}
\end{eqnarray}
$\tauvec$ is a vector to NN sites. Acting on the states $A^{\dagger}$, $B^{\dagger}$ and $C^{\dagger}$ with the Lang--Firsov Hamiltonian gives:
\begin{eqnarray}
    \tilde{H}A^{\dagger}_{\rveci}\vacu &=&\left(UA^{\dagger}_{\rveci}-t'(B^{\dagger}_{\rvec_{i}}+B^{\dagger}_{\rvec_{i}-\tauvec_{1}}+C^{\dagger}_{\rvec_{i}}+C^{\dagger}_{\rvec_{i}-\tauvec_{2}}\right)\vacu\nonumber\\
        \tilde{H}\Bdag_{\rveci}\vacu & = & \left(V\Bdag_{\rveci}-t'(\Adag_{\rveci}+\Adag_{\rveci+\tauvec_{1})}\right)\vacu\nonumber\\
                \tilde{H}\Cdag_{\rveci}\vacu & = & \left(V\Cdag_{\rveci}-t'(\Adag_{\rveci}+\Adag_{\rveci+\tauvec_{2}})\right)\vacu
\end{eqnarray}
Hence,
\begin{eqnarray}
    \tilde{H}|\Psi\rangle &= \frac{1}{N}\sum_{i}e^{i\kvec\cdot\rveci}&\left(\Adag_{\rveci}\left(aU-bt' (e^{-i\kvec\cdot\tauvec_{1}}+1)\right.\right.\nonumber\\
    & &\hspace{15mm}\left.\left.-ct' (e^{-i\kvec\cdot\tauvec_{2}}+1)\right)\right.\nonumber\\
    & & +\Bdag_{\rveci}\left(bV-at'(1+e^{i\kvec\cdot\tauvec_{1}})\right)\nonumber\\
    & & \left.+\Cdag_{\rveci}\left(cV-at'(1+e^{i\kvec\cdot\tauvec_{2}})\right)\right)\nonumber
\end{eqnarray}
Projecting onto $A, B$ and $C$ leads to secular equations,
\begin{eqnarray}
    Ea &= & aU-bt' (e^{-i\kvec\cdot\tauvec_{1}}+1)-ct' (e^{-i\kvec\cdot\tauvec_{2}}-1)\nonumber\\
    Eb & = & bV-at'(1+e^{i\kvec\cdot\tauvec_{1}})\nonumber\\
    E c & = & cV-at'(1+e^{i\kvec\cdot\tauvec_{2}})
\end{eqnarray}
which is solved by
\begin{equation}
    \left|\begin{array}{ccc}(U-E) & -t'(1+e^{-i\kvec\cdot\tauvec_{1}}) & -t'(1+e^{i\kvec\cdot\tauvec_{2}})\\
    -t'(1+e^{i\kvec\cdot\tauvec_{1}}) & (V-E) & 0 \\
    -t'(1+e^{i\kvec\cdot\tauvec_{2}}) & 0 & (V-E)\end{array}\right|=0
\end{equation}

The resulting cubic equation has three solutions, $E=V$ (an immobile intersite pair) and
\begin{equation}
    E = \frac{U+V}{2}\pm\sqrt{\frac{(U-V)^{2}}{4}+t'^{2}\left(\cos^2(\frac{k_{x}a}{2})+\cos^2(\frac{k_{y}a}{2})\right)}
\end{equation}
thus
\begin{equation}
    \frac{1}{m^{**}_{x}}=\left.\frac{1}{\hbar^2}\frac{\partial^2 E}{\partial k_{x}^{2}}\right|_{\kvec=0}
\end{equation}
\begin{equation}
    \frac{1}{m^{**}_{x}}=\frac{t^{\prime 2} a^2}{\hbar^2\sqrt{\frac{(U-V)^{2}}{4}+2t^{\prime 2}}}
\end{equation}
for a deeply bound on-site pair with $V=0$ and large, negative $U=-2W\lambda$,
\begin{equation}
    \frac{1}{m^{**}_{x}}=\frac{t^{\prime 2} a^2}{\hbar^2\sqrt{t^{2}16\lambda^2+2t^{\prime 2}}}
\end{equation}


%
%
%

%
%


%



\end{document}